\documentclass[aps,prl,reprint,groupedaddress,superscriptaddress,floatfix,citeautoscript,hyperref=pdftex]{revtex4-1}

\newcommand{\be}{\begin{equation}}
\newcommand{\ee}{\end{equation}}

\usepackage{verbatim}
\usepackage{latexsym,amssymb,float}
\usepackage{setspace}
\usepackage{graphicx}
\usepackage{dcolumn}
\usepackage{float}
\usepackage{epsfig}
\usepackage{amsmath}
\usepackage{bm}
\usepackage{lipsum}
\usepackage[colorlinks=true]{hyperref}

\begin{document}

\title{Anomalous softening of phonon-dispersion in cuprate superconductors}
\author{Saheli Sarkar}
\author{Maxence Grandadam}
\author{Catherine P\'{e}pin}

\affiliation{Institut de Physique Th\'eorique, Universit\'e Paris-Saclay, CEA, CNRS, F-91191 Gif-sur-Yvette, France.}

\begin{abstract}
A softening of phonon-dispersion has been observed experimentally in under-doped cuprate superconductors at the charge-density wave (CDW) ordering wave vector. Interestingly, the softening occurs below the superconducting (SC) transition temperature T$_{c}$, in contrast to the metallic systems, where the softening occurs usually below the CDW onset temperature T$_{\text{CDW}}$. An understanding of the `anomalous' nature of the phonon-softening and its connection to the pseudo-gap phase in under-doped cuprates remain open questions. Within a perturbative approach, we show that a complex interplay among the ubiquitous CDW, SC orders and life-time of quasi-particles associated to thermal fluctuations, can explain the anomalous phonon-softening below T$_{c}$.  Furthermore, our formalism captures different characteristics of the low temperature phonon-softening depending on material specificity.
\end{abstract}

\pacs{}

\maketitle

The `pseudo-gap' phase \cite{Alloul89,Warren89,berthier96,Marshall96,Harris97,Renner98,Ino98,
Ronning03} of the under-doped high-temperature copper-oxide based superconductors (cuprates) remains incomprehensible even after decades of research, by and large due to a complex interplay of several symmetry broken orders \cite{Fradkin:2015ch,pepin2020fluctuations}. A universally present translational symmetry broken order in the cuprates is a charge-density wave (CDW) order \cite{Hoffman02,Doiron-Leyraud07,Ghiringhelli12,wu12,Achkar12,Blackburn13a,Blackburn13b,Blanco-Canosa13,Croft14,daSilvaNeto:2014vy,matsuba07,Fujita14,machida16}. Since its discovery, the CDW order has become fundamentally important due to growing evidences of its close relation to the pseudo-gap phase, although a full knowledge about the CDW order and its relation to the pseudo-gap phase remains incomplete. One leading approach to unravel the relation, is to study the phonon-spectrum which couples to electronic degrees of freedom, thus leaving fingerprints associated to the electronic-structure.

\begin{figure}[t]
\includegraphics[width=0.6\linewidth]{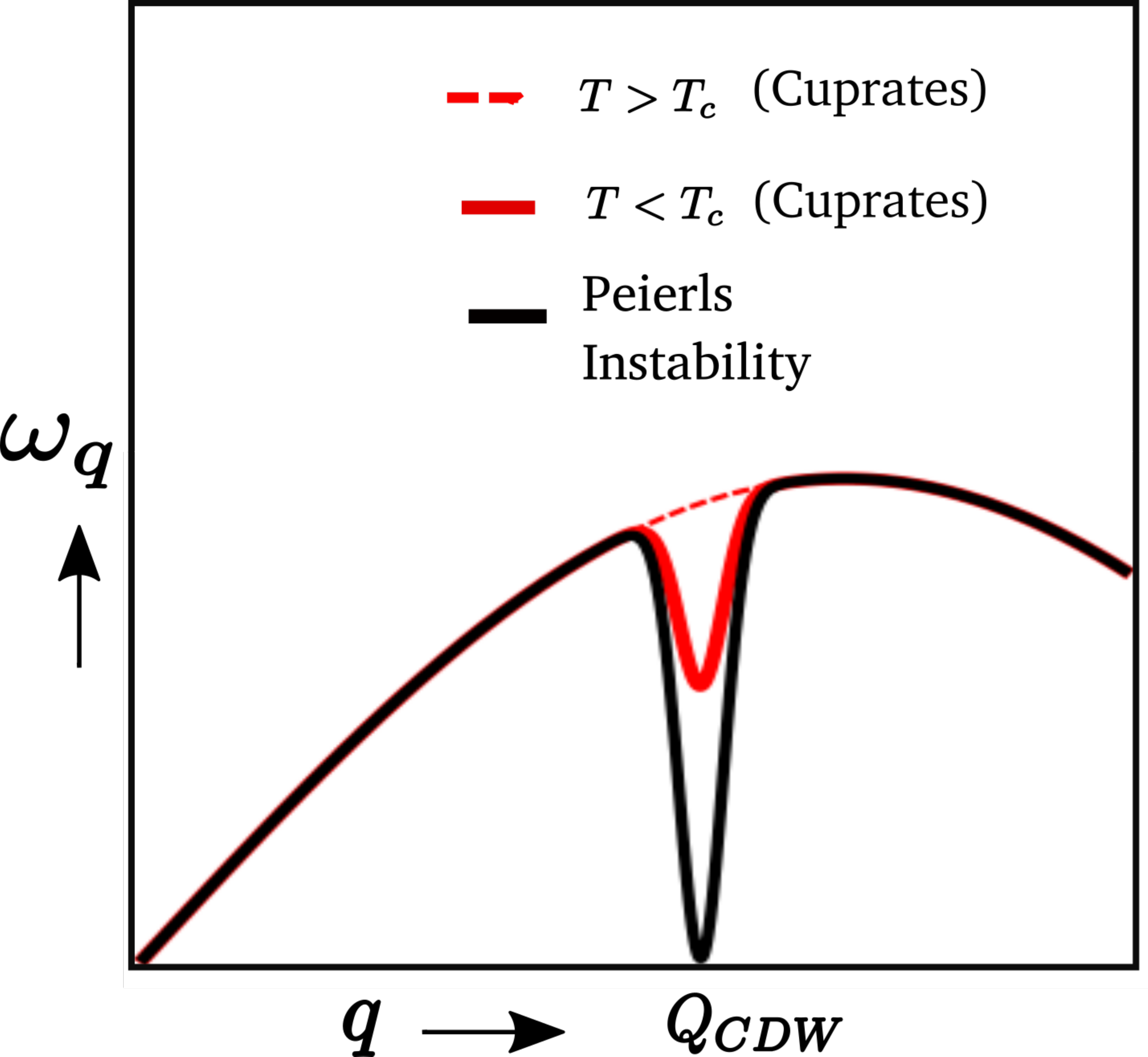}
\caption{\label{fig:kohn}Schematic representation of a full softening in metals and a partial softening in under-doped cuprates below T$_{c}$.}
\end{figure}

The phonon-spectrum has been largely studied in metallic systems, where the the charge-correlations soften the phonon-spectrum giving rise to the `Kohn-anomaly' \cite{Kohn1962}. In one dimensional metals \cite{Renker1973,Carneiro1976,Pouget1991} and in some transitional metal dichalcogenides \cite{wilson2001charge}, this softening grows towards zero [Fig.~\ref{fig:kohn}] and a full phonon-softening occurs at the CDW wave-vector (Q) below CDW ordering temperature T$_{\text{CDW}}$, reflecting the origin of CDW order in them. With a similar outlook, the phonon-spectrum has been measured even in cuprates using different experimental techniques, like inelastic x-ray scattering and inelastic neutron scattering \cite{LeTacon14,Blackburn13b,Miao18,lee2020spectroscopic,Mcqueny1999,Uchiyama04,reznik2008q,Graf08,Astuto08,baron2008first,Raichle11}. All of these experiments have observed a partial phonon-softening [Fig.~\ref{fig:kohn}] associated to Q in several cuprates, only below the superconducting transition temperature T$_{c}$, in stark contrast to the metallic systems \cite{Pouget1991,wilson2001charge,lorenzo1998neutron,Requardt02}. The occurrence of phonon-softening below T$_{c}$ is hence referred to as `anomalous' phonon-softening. 

The anomalous phonon-softening indicates a close connection between the CDW and superconductivity in under-doped cuprates. Such a connection between CDW and superconductivity have been widely discussed in various theoretical studies \cite{Efetov13,Hayward14,Wang15b,chakraborty2018phase}. Supporting evidences of this connection can also be found in several experiments \cite{Loret19,Hoffman02,Chang12,Ghiringhelli12}. Notably, a recent proposal \cite{Chakrabortyprb19}, based on the fractionalization of a pair-density wave (PDW) order \cite{edkins2019magnetic,hamidian2016detection}, advocates that for temperatures above T$_{c}$, a growing amount of fluctuations in CDW and superconductivity arising from a connection between them, can provide potential explanation to the pseudo-gap phase.

While earlier studies \cite{norman1998destruction,Norman03,Wang15b,grandadam2020electronic} discussed the role of CDW, superconductivity and associated fluctuations on the electronic-spectrum, their effect on the bosonic excitations, especially phonons, remain an outstanding question and perhaps can give a more complete understanding of the CDW orders in cuprates. In this letter, we incorporate simultaneous effects of CDW, superconductivity and thermal fluctuations on the phonon-spectrum. In our model, we mimic the fluctuations by introducing an inverse life-time of quasi-particles \cite{grandadam2020electronic,Norman:1998va} and take its temperature dependence phenomenologically \cite{Norman:1998va} based on earlier studies, which can capture various crucial aspects of the electronic spectrum in the pseudo-gap phase. We find that a strong phonon-softening occurs only below T$_{c}$, thus explaining the anomalous nature of the phonon-softening seen in experiments. Additionally, we also show that at low temperatures, different temperature dependence of the superconducting (SC) gap and inverse life-time of quasi-particle give contrasting effects on the strength of the phonon-softening.

We start with a total Hamiltonian $H_{tot}$ \cite{lee1993conductivity}, given by $H_{tot} = H_{e} + H_{ph} + H_{e-ph}$ , with,
\begin{align}
H_{e} & = \sum_{k,\sigma} \xi_{k}c_{k,\sigma}^{\dagger}c_{k,\sigma} +
 \sum_{k,\sigma}(\chi_{k}c_{k+Q,\sigma}^{\dagger}c_{k,\sigma} + h.c.) \\
 \nonumber
&+\sum_{k}(\Delta_{k}c_{k,\uparrow}^{\dagger}c_{-k,\downarrow}^{\dagger} + h.c.),\\
\nonumber
H_{ph} &= \sum_{q} \omega_{q}(b_{q}^{\dagger}b_{q}+b_{-q}^{\dagger}b_{-q}),\\
\nonumber
H_{e-ph} &= (g/\sqrt{N})\sum_{q}\sum_{k,\sigma}[c_{k+q,\sigma}^{\dagger}c_{k,\sigma}(b_{q}^{\dagger}+ b_{-q})+ h.c.],
\end{align}
where $H_{e}$ is an effective mean-field Hamiltonian with SC and CDW orders. $c^{\dagger}_{k,\sigma}(c_{k,\sigma})$ is the creation (annihilation) operator for an electron with spin $\sigma$ and momentum $k$, $\xi_{k}$ is the electronic dispersion, $\Delta_{k}$ is the SC order parameter and $\chi_{k}$ is the CDW order parameter with modulation wave-vector Q. $H_{ph}$ is the Hamiltonian for free phonons with phonon creation operator $b_{q}^{\dagger}$ for wave-vector q and frequency $\omega_{q}$. $H_{e-ph}$ is the Hamiltonian describing electron-phonon interaction with strength $g$ and $N$ is the number of lattice sites in the system. The Green's function corresponding to $H_{e}$ is given by $\hat G^{-1}(i\omega_{n},k) = (i \omega_{n} -\hat H_{e} )$ and has a matrix form in the extended Nambu basis $\Psi^{\dagger}_{k}  = \left(c^{\dagger}_{k,\uparrow},c_{-k,\downarrow},c^{\dagger}_{k+Q,\uparrow},c_{-k-Q,\downarrow}\right)$ which is given by,
\begin{align}\label{eq_greenmat}
\tiny
G^{-1}&=
\begin{pmatrix}
i\omega_{n}-\xi_{k} & -\Delta_{k} & -\chi_{k} &0 \\
-\Delta^{*}_{k} & i\omega_{n}+\xi_{k} & 0 &\chi_{k}\\
-\chi^{*}_{k} &0 & i\omega_{n} - \xi_{k+Q} & -\Delta_{k+Q}\\
0 & \chi^{*}_{k} &-\Delta^{*}_{k+Q} & i\omega_{n}+\xi_{k+Q}
\end{pmatrix},
\end{align}
where $\omega_{n}$ is the Matsubara frequency. We use a band-structure for a prototype cuprate system \cite{Berthod17} [see supplementary materials (SM) \cite{SuppInf}]. We consider a d-wave symmetric SC gap, given by $\Delta_{k} = (\Delta_{max}/2)[\cos(k_{x})-\cos(k_{y})]$, where $\Delta_{max}$ denotes the maximum gap. Following several theoretical studies \cite{Wang:2014fr,Chowdhury:2014cp,Chakrabortyprb19} and experimental evidences \cite{daSilvaNeto:2014vy,Comin14}, we consider a CDW order parameter with Q given by the axial wave-vector connecting two neighboring `hot-spots', the points on Fermi-surface which intersect the magnetic-brillouin zone boundary \cite{Efetov13}. Moreover, the CDW gap is taken to have a maximum ($\chi_{max}$) at the hot-spots, falling off exponentially away from the hot-spots \cite{Chakrabortyprb19}. 

The modified electronic spectrum in the presence of SC and CDW orders will re-normalize the free phonon propagator, $D_{0}(z,q)= 2 \omega_{q}/(z^{2}-\omega^{2}_{q})$. To analyze this, we begin by writing the imaginary time ($\tau$) phonon propagators in matrix form in the ordered phase. The corresponding matrix elements are given by $D_{m,n}(q,\tau) = -\langle \mathcal{T} \phi_{q+mQ}(\tau)\phi^{\dagger}_{q+nQ}(0)\rangle$, where $\mathcal{T}$ is the time-ordering operator \cite{lee1993conductivity}, $\phi_{q}$ is the phonon field operator given by $b_{q}^{\dagger} + b_{-q}$ and $m,n = \pm$. Noting that $D_{++} \equiv D_{--}:=D_{1}(z,q)$ and $D_{+-} \equiv D_{-+}:=D_{2}(z,q)$, within a perturbative treatment of electron-phonon interaction, we evaluate the re-normalized phonon propagators $D_{1}$ and $D_{2}$ by using Dyson equations
\begin{align}\label{eq:Dyson}
\tiny
D_{1}(z,q) &= D_{0}(z,q+Q)\bigg[1 + \Pi_{1}(z,q)D_{1}(z,q)+\\
\nonumber
&\Pi_{2}(z,q)D_{1}(z,q) +\Pi_{3}(z,q)D_{2}(z,q)+\\
\nonumber
&\Pi_{4}(z,q)D_{2}(z,q)\bigg],\\
\nonumber
 D_{2}(z,q) & =  D_{0}(z,q-Q)\bigg[\Pi_{1}(z,q)D_{2}(z,q)+\Pi_{2}(z,q)D_{2}(z,q) \\
\nonumber
&+\Pi_{3}(z,q)D_{1}(z,q)+ \Pi_{4}(z,q)D_{1}(z,q)\bigg],
\end{align}
\begin{figure}[t]
\includegraphics[width=1.0\linewidth]{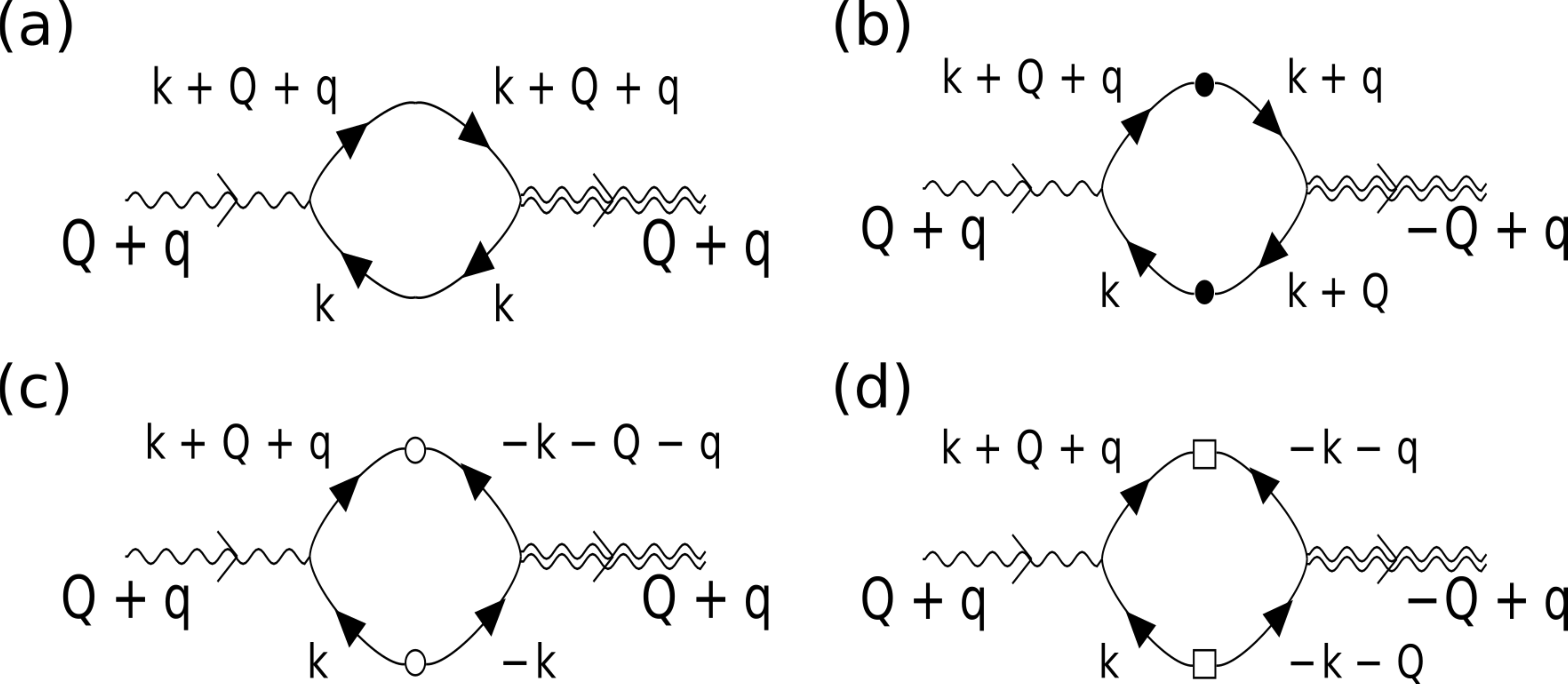}
\caption{\label{fig2} (a), (b), (c) and (d) represent the Feynman diagrams for the terms in the Dyson equations [Eq.~\eqref{eq:Dyson}] involving the self-energies $\Pi_{1}$, $\Pi_{2}$, $\Pi_{3}$ and $\Pi_{4}$ respectively in the presence of CDW and SC orders.}
\end{figure}
where, $\Pi_{1,2,3,4}(z,q)$ represent the phonon self-energies. The leading contributions to the Dyson equations [Eqs.~\eqref{eq:Dyson}] are shown in Fig.~\ref{fig2}. Explicit expressions for  $\Pi_{1,2,3,4}(z,q)$ are presented in the SM \cite{SuppInf}.
\\
We obtain the new modes for phonon in the ordered phase by decoupling Eq.~\eqref{eq:Dyson}, with the definition $D_{\pm}(z,q) = D_{1}(z,q) \pm D_{2}(z,q)$ and then solving $D_{\pm}(z,q)$ with the assumption that $\omega_{Q\pm q} \approx \omega_{Q}$ for small q. Finally, plugging in $D_{0}(z,q)$, we obtain the solutions as,
\begin{align}\label{Eq:Modephonon}
D_{\pm}(z,q) = \frac{2\omega_{Q}}{z^{2}-\omega_{Q}^{2}-2\omega_{Q}\Pi_{\pm}(z,q)},
\end{align}
where $\Pi_{+} = \Pi_{1}+\Pi_{2}+\Pi_{3}+\Pi_{4} $ and $\Pi_{-} = \Pi_{1}+\Pi_{2}-\Pi_{3}-\Pi_{4}$.
The dispersion of the new phonon modes correspond to the values of $z$, for which denominator of Eq.~\eqref{Eq:Modephonon} vanishes. Subsequently, taking only $q$ dependence in $\Pi$, the frequency for each mode is given by
\begin{align}\label{eq:dispersion}
\Omega_{\pm}^{2}(q)=\omega_{Q}^{2}+2\omega_{Q}\Pi_{\pm}(q).
\end{align}
These two new phonon modes in Eq.~\eqref{eq:dispersion} with frequency $\Omega_{\pm}$ signify branching of the free phonon near Q due to presence of CDW and SC orders. We find that the split between $\Omega_{\pm}$ is proportional to the magnitude of the CDW order. Also, we only plot $\Pi_{\pm}$ as a function of $\tilde{q} = q-Q$, as the modes $\Omega_{\pm}(q)$ can be easily identified from the corresponding $\Pi_{\pm}$ in Eq.~\eqref{eq:dispersion}. For depicting the strength of the phonon-softening, we look at $\Pi_{\pm}(\tilde{q})$ after subtracting $\Pi_{\pm}(\tilde{q} = -1)$.  In Fig.~\ref{fig:phonon}, we observe that $\Pi_{\pm}(\tilde{q})$ decreases strongly within a finite range around $\tilde{q} =0$, with a minimum at $\tilde{q} =0$, readily suggesting a softening of phonon-frequency around Q. We also observe that, away from $\tilde{q}= 0$, $\Pi_{\pm}(\tilde{q})$ goes towards zero, implying a suppression of phonon-softening away from Q. This suggests that the effect of CDW and SC orders on the phonon are maximum at Q, and diminishes away from it. Additionally, we notice that the suppression of $\Pi_{-}$ is more than the suppression of $\Pi_{+}$ and the $\tilde{q}$ dependence of $\Pi_{\pm}$ are extremely similar to each other. Hence, for a simpler presentation, in the rest of the paper, we only plot $\Pi_{-}$ with $\tilde{q}$ [relabeled as $\Pi(\tilde{q})$].
\begin{figure}[t]
\includegraphics[width=0.75\linewidth]{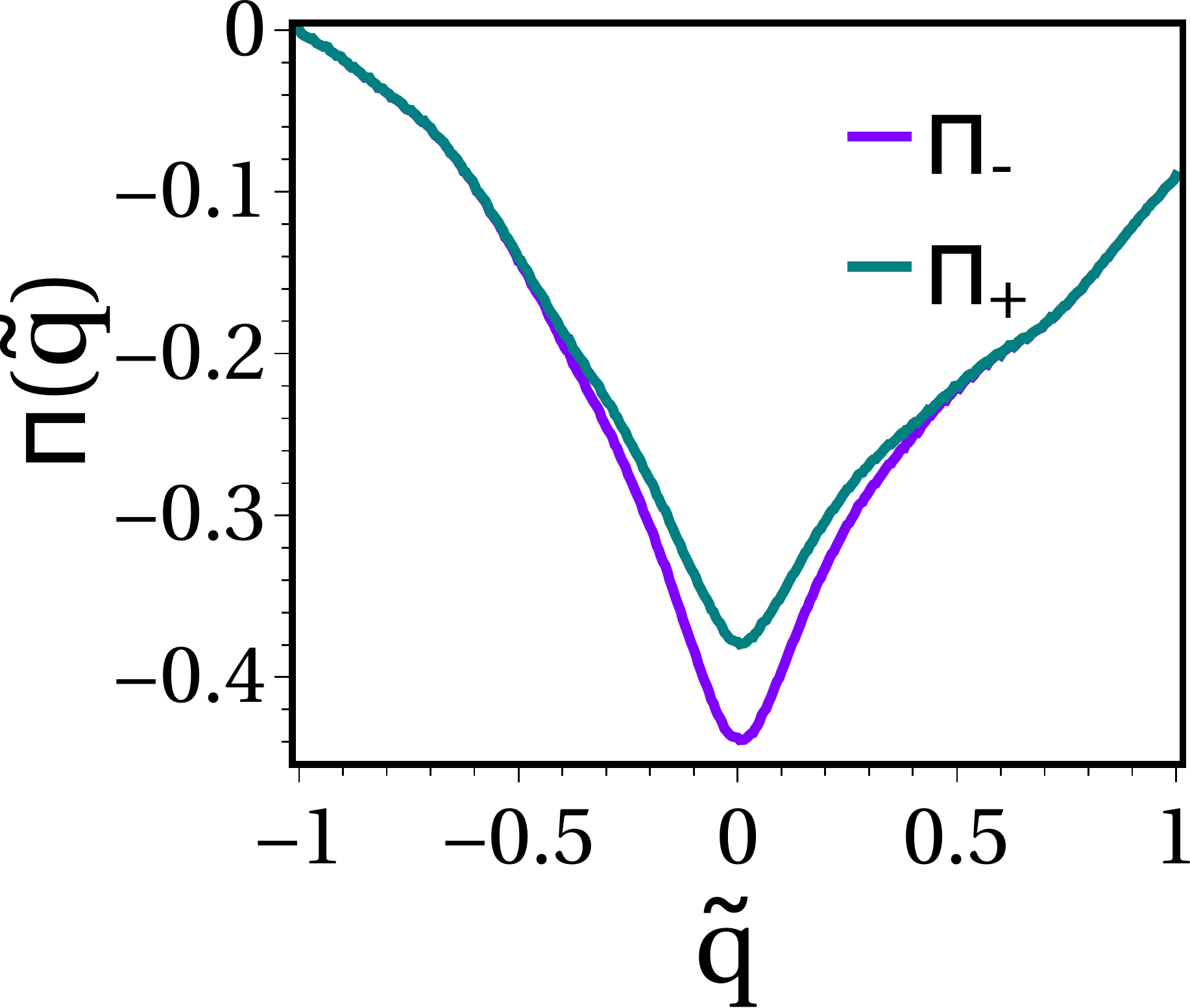}
\caption{\label{fig:phonon} Plots of the self-energy $\Pi_{\pm}$ as a function of $\tilde{q} = q-Q$ corresponding to the two re-normalized phonon modes $\Omega_{\pm}$ in the presence of $\chi_{max} = 0.05$ and $\Delta_{max} = 0.05$. Both $\Pi_{\pm}$ exhibit a depletion around $\tilde{q} = 0$, implying a softening in the phonon-dispersion of the two new modes $\Omega_{\pm}$ around Q.}
\end{figure}
\\
So far, we obtain a phonon-softening in the presence of SC and CDW orders. However, to address the anomalous phonon-softening in cuprates, we need to include fluctuation related effects, which are major constituents governing the phase diagram of these systems. For example, such fluctuations can lead to quasi-particle scattering which are known to have vital roles in Fermi-arc related physics of the pseudo-gap phase \cite{Norman:1995dd,NormanKanigel07,dalla2015exploring}. The strength of the scattering depends on temperature; while it can be large at high temperatures, a sudden reduction occurs below T$_{c}$, which can be attributed to a fractionalization of a PDW \cite{Chakrabortyprb19}. To give an idea, the proposal of fractionalization of a PDW order suggests that the fluctuation of a U(1) gauge field gives a constraint connecting SC and CDW. As a result, fractionalization of PDW occurs at an energy scale associated to the pseudo-gap temperature T*, consequently fluctuations largely increase in the system. However, below T$_{c}$, the fluctuations quench, thus yielding a global phase coherence of CDW and SC orders and increasing the life-time of quasi-particles.
 
In order to study the evolution of the phonon-softening with temperature, we incorporate a finite inverse life-time of quasi-particles, given by $\Gamma$, pertinent to the fluctuation related effects in the system. The self-energy in Matsubara frequency due to $\Gamma$ can be written as $\Sigma = i\Gamma \textit{sgn}(\omega_{n})$ and the Green's function in Eq.~\eqref{eq_greenmat} will transform as
\begin{align}\label{eq:lifetime}
G_{i,j}^{-1}(i \omega_{n},k) \rightarrow G_{i,j}^{-1}(i \omega_{n} + \Sigma,k).
\end{align}
In the presence of $\Gamma$, the phonon-dispersion will be modified by the real part of $\Pi(\tilde{q})$, again relabeled as $\Pi(\tilde{q})$. Detailed calculations are presented in the SM \cite{SuppInf}.

\begin{figure}[t]
\includegraphics[width=0.99\linewidth]{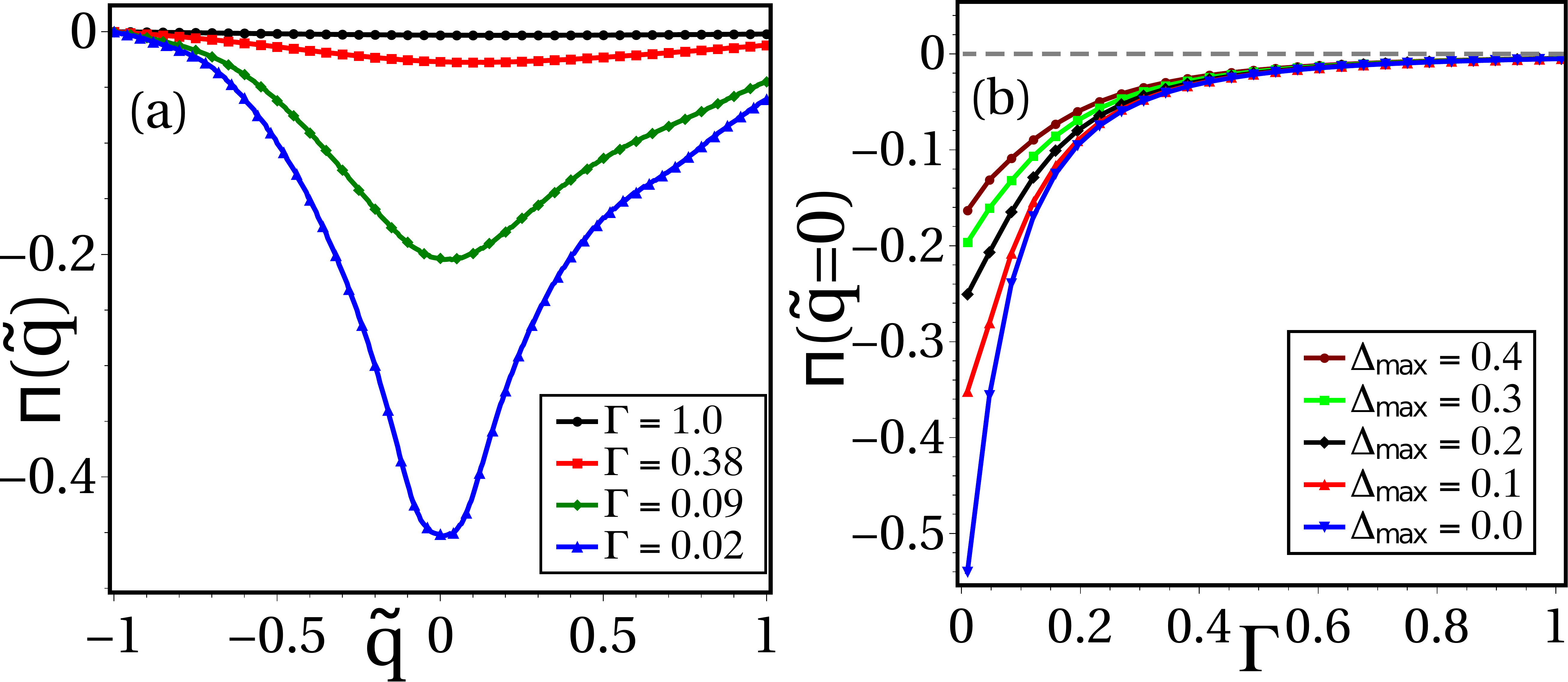}
\caption{\label{fig4}~(a) The variation of $\Pi(\tilde{q})$ with $\tilde{q}$ for four different values of $\Gamma$ with $\chi_{max} = 0.2$ and $\Delta_{max} = 0$. The plots portray a suppression in phonon-softening with increase in $\Gamma$. (b) Plots of $\Pi(\tilde{q}=0)$ with variation in $\Gamma$ for five different values of  $\Delta_{max}$ with $\chi_{max} = 0.2$. The plots manifest a suppression in phonon-softening with an increase in $\Delta_{max}$. The effect of $\Delta_{max}$  is strongest for low $\Gamma$, and weakest for high $\Gamma$.}
\end{figure}

To understand the collective effect of the SC gap and $\Gamma$ on the phonon-softening, it is important to disentangle the role played by $\Gamma$ and the SC gap. Therefore, we start by studying the effect of $\Gamma$ taking $\Delta_{max} = 0$. Fig.~\ref{fig4}(a) shows the variation of $\Pi(\tilde{q})$ as a function of $\tilde{q}$ for four different $\Gamma$ with $\chi_{max} = 0.2$. We notice that for very small value of $\Gamma = 0.02$, there is a significantly strong phonon-softening around $\tilde{q}=0$. With increasing $\Gamma$, the phonon-softening starts to reduce and for a very large $\Gamma = 1.0$, the phonon-softening gets almost fully suppressed. We also observe that the phonon-softening at $\tilde{q} =0$ is most strongly affected by $\Gamma$. Therefore, for rest of the analysis, we will concentrate on $\Pi$ at $\tilde{q} =0$ to quantify the phonon-softening.
\\
Now, we inspect the role of the SC order and the interplay between superconductivity and $\Gamma$. In Fig.~\ref{fig4}(b), we plot the variation of $\Pi(\tilde{q}=0)$ with $\Gamma$, for five different $\Delta_{max}$ taking $\chi_{max} = 0.2$. We notice that $\Delta_{max}$ has a prominent effect when $\Gamma$ is very small, as can be seen from the change in $\Pi(\tilde{q}=0)$ around $\Gamma \sim 0.05$. In this regime, $\Delta_{max}$ weakens the softening of phonon. Similar effect on phonons in the SC phase has been indicated in conventional s-wave superconductors \cite{Axe1973,reznik2012phonon}. With increasing $\Gamma$, for example around $\Gamma \sim 0.3$,  the effect of $\Delta_{max}$ becomes less significant. Finally, for very large $\Gamma\simeq 1.0$, changing $\Delta_{max}$ has almost no effect. These results highlight two crucial points. First, both superconductivity and $\Gamma$ suppress the phonon-softening. Second, the role of $\Delta_{max}$ is prominent at low $\Gamma$, while negligible for large $\Gamma$.

We have seen that the introduction of superconductivity suppresses the phonon-softening, while experiments observe a seemingly opposite characteristic of enhancement of phonon-softening below T$_{c}$. At this point, we should also notice that $\Gamma$ suppresses the phonon-softening, as shown in Fig.~\ref{fig4}(a). Moreover, $\Gamma$ is expected to increase with temperature due to increase in fluctuations, whereas $\Delta_{max}$ is expected to decrease with temperature, for example in a simple BCS type scenario. Thus, they behave in opposite manner with temperature.  
\begin{figure}[t]
\includegraphics[width=0.98\linewidth]{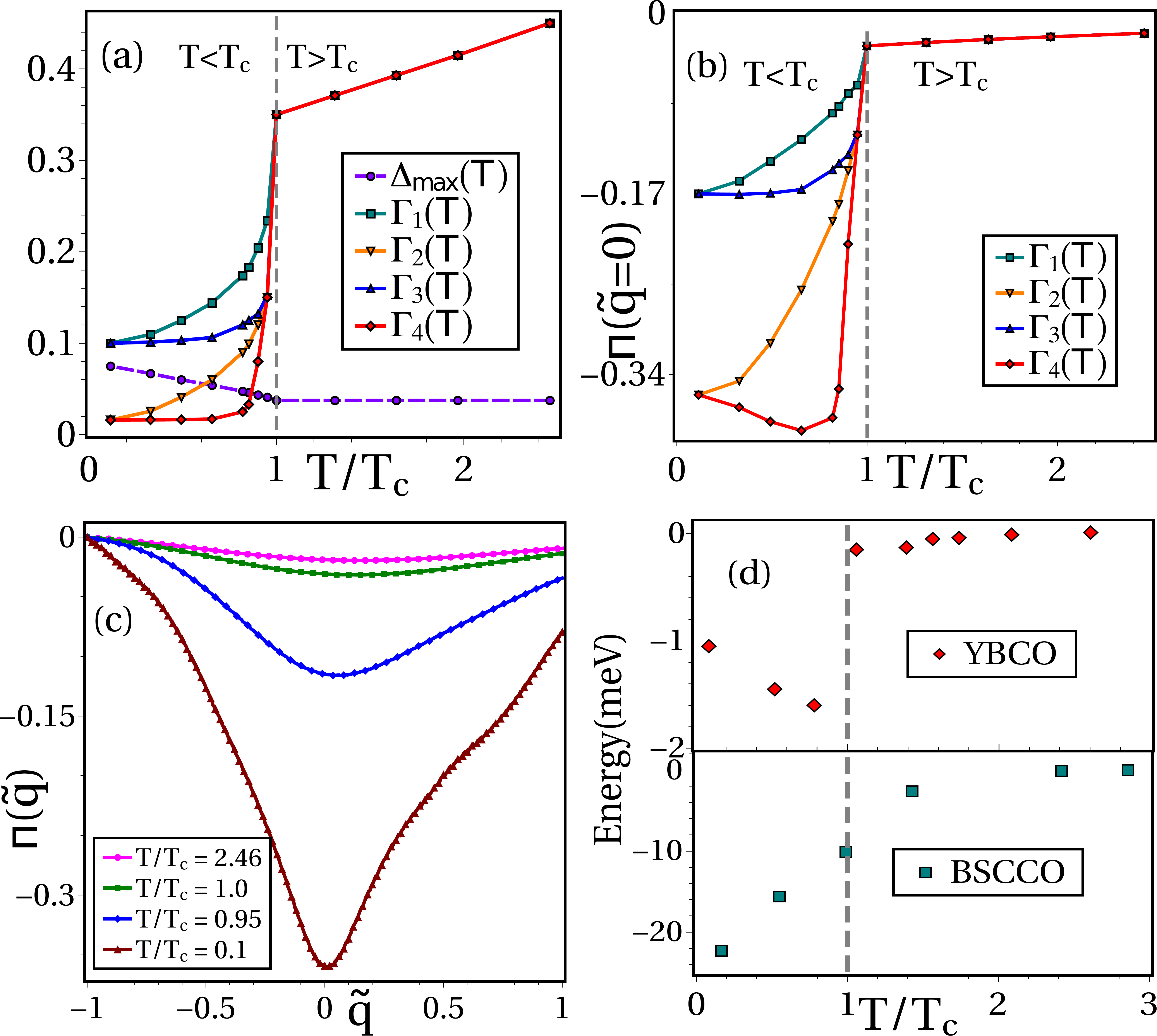}
\caption{\label{Fig:5} (a) Different sets of T-dependence for inverse life-time of quasiparticles denoted by $\Gamma_{1}$, $\Gamma_{2}$, $\Gamma_{3}$ and $\Gamma_{4}$. The T-dependence of the SC gap is denoted by $\Delta_{max}$(T). In all cases, $\chi_{max} = \Delta_{max}$. (b) The T-dependence of $\Pi(\tilde{q}=0)$ for different parameter sets in (a). A large negative value of $\Pi(\tilde{q}=0)$ in the regime T $ \lesssim$ T$_{c}$ implies a strong enhancement of phonon-softening, while $\Pi(\tilde{q} =0) \rightarrow 0$ implies a strong suppression in phonon-softening in the regime T $>$ T$_{c}$. (c) The variation of $\Pi(\tilde{q})$ with $\tilde{q}$ at four different temperatures for parameter set $\Gamma_{4}$ and $\Delta_{max}$(T) shown in (a). (d) Schematic representation of the experimental results of phonon-softening at CDW wave-vector for YBCO and BSCCO, adopted from Refs.~\cite{LeTacon14,lee2020spectroscopic}.}
\end{figure}

We consider temperature (T) dependence phenomenologically in $\Delta_{max}$ and $\Gamma$, similar to the T dependence used in explaining spectral function in ARPES experiments \cite{Norman:1998va}. The T-dependence of $\Delta_{max}$ and $\Gamma$ are shown in Fig.~\ref{Fig:5}(a). Below T$_{c}$, $\Delta_{max}$ decreases with T, whereas remains approximately constant above T$_{c}$. Moreover following indications from Raman spectroscopy \cite{Loret19}, $\chi_{max}$ is taken to be equal to $\Delta_{max}$. To illustrate how different T-dependence of $\Gamma$ and $\Delta_{max}$ can give different features in phonon-softening, we use four different types of T-dependence for $\Gamma$, denoted by $\Gamma_{1}, \Gamma_{2}, \Gamma_{3}$ and $\Gamma_{4}$ in Fig.~\ref{Fig:5}(a). Note that, they differ in magnitudes compared to $\Delta_{max}$. In all these cases, $\Gamma$ reduces significantly below T$_{c}$,  with the strongest fall in $\Gamma_{4}$ and the weakest fall in $\Gamma_{1}$, but still remains finite even in the limit T$\rightarrow 0$ \cite{chubukov2007gapless}. Moreover, we considered in all the cases, a linear T-dependence for $\Gamma$ for T $ >$ T$_{c}$, as suggested in some earlier works  \cite{kanigel2006evolution,Varma89}. 

In Fig.~\ref{Fig:5}(b), we plot $\Pi(\tilde{q}=0)$ for the parameters in Fig.~\ref{Fig:5}(a). We start by closely inspecting the $\Gamma_{4}$ case in Fig.~\ref{Fig:5}(b). We observe that the values of $\Pi(\tilde{q}=0)$ are close to zero for high temperatures (T $\gg$ T$_{c}$), implying that the phonon-softening is strongly suppressed. Remarkably, we observe that for temperatures T $ \lesssim$ T$_{c}$, the values of $\Pi(\tilde{q}=0)$ reduce sharply towards more negative values, which suggest that the phonon-softening enhances strongly. But surprisingly, towards further lower temperatures below T$_{c}$, $\Pi(\tilde{q}=0)$ enhances, which implies a suppression in phonon-softening. However, the phonon-softening below T$_{c}$ always remains stronger as compared to T $>$ T$_{c}$. Very similar features have been observed in YBa$_{2}$Cu$_{3}$O$_{6+y}$ (YBCO) \cite{LeTacon14}, as shown schematically in Fig.~\ref{Fig:5}(d). In Fig.~\ref{Fig:5}(c), we present the the full $\tilde{q}$ dependence of $\Pi$ at four different temperatures for the case $\Gamma_{4}$. We observe that away from $\tilde{q} = 0$, phonon-softening is less sensitive to the variation of temperature. Similar feature has been found in experiments \cite{LeTacon14,lee2020spectroscopic}.

Next, we closely investigate the $\Gamma_{1}$ case in Fig.~\ref{Fig:5}(b) for T $\lesssim$ T$_{c}$. Very interestingly, the features for T $\lesssim$ T$_{c}$ possess marked differences from $\Gamma_{4}$ case. We notice a smoother enhancement in phonon-softening just below T$_{c}$ (T $\sim$ T$_{c}$), while the enhancement is more rapid and sharper for $\Gamma_{4}$ case. In particular, towards lower temperatures (T $\rightarrow$ 0), a further enhancement in phonon-softening can be noticed in contrast to the suppression observed for $\Gamma_{4}$. Analogous features in phonon-softening have been also observed in Bi$_{2}$Sr$_{2}$CaCu$_{2}$O$_{8+y}$ (BSCCO) \cite{lee2020spectroscopic}, schematically presented in Fig.~\ref{Fig:5}(d). To demonstrate the different features in phonon-softening resulting from an intricate interplay between SC gap and $\Gamma$ below T$_{c}$, we plot results for two more cases $\Gamma_{2}$ and $\Gamma_{3}$, shown in Fig.~\ref{Fig:5}(b). Below T$_{c}$, for $\Gamma_{2}$, phonon-softening sharply enhances than for $\Gamma_{3}$ as T $\rightarrow$ 0.
 
In summary, within a mean-field description of superconductivity and charge-density wave (CDW), describing under-doped cuprates, we obtained a softening of the phonon-dispersion associated to the CDW wave-vector (Q). The crucial finding of our work is that reduced amount of fluctuations in both CDW and superconducting (SC) orders below T$_{c}$, can successfully describe the `anomalous' phonon-softening. A reduction in the fluctuations below T$_{c}$ can be motivated from a recent proposal based on fractionalization of a PDW order \cite{Chakrabortyprb19}. Moreover, we also found that the features of phonon-softening at low temperatures depend on an intricate interplay between SC order and fluctuations. In this work, we considered the strength of electron-phonon coupling to be momentum (k)-independent. However, the formalism in this work, can easily be extended to include k-dependent electron-phonon coupling. We expect, in such a scenario, the phonon-softening will still occur at Q only below T$_{c}$, but the softening will have a different wave-vector dependence around Q. We believe our results can find applications in many two-dimensional materials where an interplay between CDW and SC orders plays an important role and thus opening much broader prospects of our work.

We acknowledge A. Banerjee and Y. Sidis for valuable discussions. This work has received financial support from the ERC, under grant agreement AdG-694651-CHAMPAGNE.

\bibliographystyle{apsrev4-1}
\bibliography{Cuprates}

\begin{thebibliography}{67}%
\makeatletter
\providecommand \@ifxundefined [1]{%
 \@ifx{#1\undefined}
}%
\providecommand \@ifnum [1]{%
 \ifnum #1\expandafter \@firstoftwo
 \else \expandafter \@secondoftwo
 \fi
}%
\providecommand \@ifx [1]{%
 \ifx #1\expandafter \@firstoftwo
 \else \expandafter \@secondoftwo
 \fi
}%
\providecommand \natexlab [1]{#1}%
\providecommand \enquote  [1]{``#1''}%
\providecommand \bibnamefont  [1]{#1}%
\providecommand \bibfnamefont [1]{#1}%
\providecommand \citenamefont [1]{#1}%
\providecommand \href@noop [0]{\@secondoftwo}%
\providecommand \href [0]{\begingroup \@sanitize@url \@href}%
\providecommand \@href[1]{\@@startlink{#1}\@@href}%
\providecommand \@@href[1]{\endgroup#1\@@endlink}%
\providecommand \@sanitize@url [0]{\catcode `\\12\catcode `\$12\catcode
  `\&12\catcode `\#12\catcode `\^12\catcode `\_12\catcode `\%12\relax}%
\providecommand \@@startlink[1]{}%
\providecommand \@@endlink[0]{}%
\providecommand \url  [0]{\begingroup\@sanitize@url \@url }%
\providecommand \@url [1]{\endgroup\@href {#1}{\urlprefix }}%
\providecommand \urlprefix  [0]{URL }%
\providecommand \Eprint [0]{\href }%
\providecommand \doibase [0]{http://dx.doi.org/}%
\providecommand \selectlanguage [0]{\@gobble}%
\providecommand \bibinfo  [0]{\@secondoftwo}%
\providecommand \bibfield  [0]{\@secondoftwo}%
\providecommand \translation [1]{[#1]}%
\providecommand \BibitemOpen [0]{}%
\providecommand \bibitemStop [0]{}%
\providecommand \bibitemNoStop [0]{.\EOS\space}%
\providecommand \EOS [0]{\spacefactor3000\relax}%
\providecommand \BibitemShut  [1]{\csname bibitem#1\endcsname}%
\let\auto@bib@innerbib\@empty
\bibitem [{\citenamefont {Alloul}\ \emph {et~al.}(1989)\citenamefont {Alloul},
  \citenamefont {Ohno},\ and\ \citenamefont {Mendels}}]{Alloul89}%
  \BibitemOpen
  \bibfield  {author} {\bibinfo {author} {\bibfnamefont {H.}~\bibnamefont
  {Alloul}}, \bibinfo {author} {\bibfnamefont {T.}~\bibnamefont {Ohno}}, \ and\
  \bibinfo {author} {\bibfnamefont {P.}~\bibnamefont {Mendels}},\ }\href
  {\doibase 10.1103/PhysRevLett.63.1700} {\bibfield  {journal} {\bibinfo
  {journal} {Phys. Rev. Lett.}\ }\textbf {\bibinfo {volume} {63}},\ \bibinfo
  {pages} {1700} (\bibinfo {year} {1989})}\BibitemShut {NoStop}%
\bibitem [{\citenamefont {Warren}\ \emph {et~al.}(1989)\citenamefont {Warren},
  \citenamefont {Walstedt}, \citenamefont {Brennert}, \citenamefont {Cava},
  \citenamefont {Tycko}, \citenamefont {Bell},\ and\ \citenamefont
  {Dabbagh}}]{Warren89}%
  \BibitemOpen
  \bibfield  {author} {\bibinfo {author} {\bibfnamefont {W.~W.}\ \bibnamefont
  {Warren}}, \bibinfo {author} {\bibfnamefont {R.~E.}\ \bibnamefont
  {Walstedt}}, \bibinfo {author} {\bibfnamefont {G.~F.}\ \bibnamefont
  {Brennert}}, \bibinfo {author} {\bibfnamefont {R.~J.}\ \bibnamefont {Cava}},
  \bibinfo {author} {\bibfnamefont {R.}~\bibnamefont {Tycko}}, \bibinfo
  {author} {\bibfnamefont {R.~F.}\ \bibnamefont {Bell}}, \ and\ \bibinfo
  {author} {\bibfnamefont {G.}~\bibnamefont {Dabbagh}},\ }\href {\doibase
  10.1103/PhysRevLett.62.1193} {\bibfield  {journal} {\bibinfo  {journal}
  {Phys. Rev. Lett.}\ }\textbf {\bibinfo {volume} {62}},\ \bibinfo {pages}
  {1193} (\bibinfo {year} {1989})}\BibitemShut {NoStop}%
\bibitem [{\citenamefont {Berthier}\ \emph {et~al.}(1996)\citenamefont
  {Berthier}, \citenamefont {Julien}, \citenamefont {Horvati{\'c}},\ and\
  \citenamefont {Berthier}}]{berthier96}%
  \BibitemOpen
  \bibfield  {author} {\bibinfo {author} {\bibfnamefont {C.}~\bibnamefont
  {Berthier}}, \bibinfo {author} {\bibfnamefont {M.}~\bibnamefont {Julien}},
  \bibinfo {author} {\bibfnamefont {M.}~\bibnamefont {Horvati{\'c}}}, \ and\
  \bibinfo {author} {\bibfnamefont {Y.}~\bibnamefont {Berthier}},\ }\href@noop
  {} {\bibfield  {journal} {\bibinfo  {journal} {Journal de Physique I}\
  }\textbf {\bibinfo {volume} {6}},\ \bibinfo {pages} {2205} (\bibinfo {year}
  {1996})}\BibitemShut {NoStop}%
\bibitem [{\citenamefont {Marshall}\ \emph {et~al.}(1996)\citenamefont
  {Marshall}, \citenamefont {Dessau}, \citenamefont {Loeser}, \citenamefont
  {Park}, \citenamefont {Matsuura}, \citenamefont {Eckstein}, \citenamefont
  {Bozovic}, \citenamefont {Fournier}, \citenamefont {Kapitulnik},
  \citenamefont {Spicer},\ and\ \citenamefont {Shen}}]{Marshall96}%
  \BibitemOpen
  \bibfield  {author} {\bibinfo {author} {\bibfnamefont {D.~S.}\ \bibnamefont
  {Marshall}}, \bibinfo {author} {\bibfnamefont {D.~S.}\ \bibnamefont
  {Dessau}}, \bibinfo {author} {\bibfnamefont {A.~G.}\ \bibnamefont {Loeser}},
  \bibinfo {author} {\bibfnamefont {C.-H.}\ \bibnamefont {Park}}, \bibinfo
  {author} {\bibfnamefont {A.~Y.}\ \bibnamefont {Matsuura}}, \bibinfo {author}
  {\bibfnamefont {J.~N.}\ \bibnamefont {Eckstein}}, \bibinfo {author}
  {\bibfnamefont {I.}~\bibnamefont {Bozovic}}, \bibinfo {author} {\bibfnamefont
  {P.}~\bibnamefont {Fournier}}, \bibinfo {author} {\bibfnamefont
  {A.}~\bibnamefont {Kapitulnik}}, \bibinfo {author} {\bibfnamefont {W.~E.}\
  \bibnamefont {Spicer}}, \ and\ \bibinfo {author} {\bibfnamefont {Z.-X.}\
  \bibnamefont {Shen}},\ }\href {\doibase 10.1103/PhysRevLett.76.4841}
  {\bibfield  {journal} {\bibinfo  {journal} {Phys. Rev. Lett.}\ }\textbf
  {\bibinfo {volume} {76}},\ \bibinfo {pages} {4841} (\bibinfo {year}
  {1996})}\BibitemShut {NoStop}%
\bibitem [{\citenamefont {Harris}\ \emph {et~al.}(1997)\citenamefont {Harris},
  \citenamefont {White}, \citenamefont {Shen}, \citenamefont {Ikeda},
  \citenamefont {Yoshizaki}, \citenamefont {Eisaki}, \citenamefont {Uchida},
  \citenamefont {Si}, \citenamefont {Xiong}, \citenamefont {Zhao},\ and\
  \citenamefont {Dessau}}]{Harris97}%
  \BibitemOpen
  \bibfield  {author} {\bibinfo {author} {\bibfnamefont {J.~M.}\ \bibnamefont
  {Harris}}, \bibinfo {author} {\bibfnamefont {P.~J.}\ \bibnamefont {White}},
  \bibinfo {author} {\bibfnamefont {Z.-X.}\ \bibnamefont {Shen}}, \bibinfo
  {author} {\bibfnamefont {H.}~\bibnamefont {Ikeda}}, \bibinfo {author}
  {\bibfnamefont {R.}~\bibnamefont {Yoshizaki}}, \bibinfo {author}
  {\bibfnamefont {H.}~\bibnamefont {Eisaki}}, \bibinfo {author} {\bibfnamefont
  {S.}~\bibnamefont {Uchida}}, \bibinfo {author} {\bibfnamefont {W.~D.}\
  \bibnamefont {Si}}, \bibinfo {author} {\bibfnamefont {J.~W.}\ \bibnamefont
  {Xiong}}, \bibinfo {author} {\bibfnamefont {Z.-X.}\ \bibnamefont {Zhao}}, \
  and\ \bibinfo {author} {\bibfnamefont {D.~S.}\ \bibnamefont {Dessau}},\
  }\href {\doibase 10.1103/PhysRevLett.79.143} {\bibfield  {journal} {\bibinfo
  {journal} {Phys. Rev. Lett.}\ }\textbf {\bibinfo {volume} {79}},\ \bibinfo
  {pages} {143} (\bibinfo {year} {1997})}\BibitemShut {NoStop}%
\bibitem [{\citenamefont {Renner}\ \emph {et~al.}(1998)\citenamefont {Renner},
  \citenamefont {Revaz}, \citenamefont {Genoud}, \citenamefont {Kadowaki},\
  and\ \citenamefont {Fischer}}]{Renner98}%
  \BibitemOpen
  \bibfield  {author} {\bibinfo {author} {\bibfnamefont {C.}~\bibnamefont
  {Renner}}, \bibinfo {author} {\bibfnamefont {B.}~\bibnamefont {Revaz}},
  \bibinfo {author} {\bibfnamefont {J.-Y.}\ \bibnamefont {Genoud}}, \bibinfo
  {author} {\bibfnamefont {K.}~\bibnamefont {Kadowaki}}, \ and\ \bibinfo
  {author} {\bibfnamefont {O.}~\bibnamefont {Fischer}},\ }\href {\doibase
  10.1103/PhysRevLett.80.149} {\bibfield  {journal} {\bibinfo  {journal} {Phys.
  Rev. Lett.}\ }\textbf {\bibinfo {volume} {80}},\ \bibinfo {pages} {149}
  (\bibinfo {year} {1998})}\BibitemShut {NoStop}%
\bibitem [{\citenamefont {Ino}\ \emph {et~al.}(1998)\citenamefont {Ino},
  \citenamefont {Mizokawa}, \citenamefont {Kobayashi}, \citenamefont
  {Fujimori}, \citenamefont {Sasagawa}, \citenamefont {Kimura}, \citenamefont
  {Kishio}, \citenamefont {Tamasaku}, \citenamefont {Eisaki},\ and\
  \citenamefont {Uchida}}]{Ino98}%
  \BibitemOpen
  \bibfield  {author} {\bibinfo {author} {\bibfnamefont {A.}~\bibnamefont
  {Ino}}, \bibinfo {author} {\bibfnamefont {T.}~\bibnamefont {Mizokawa}},
  \bibinfo {author} {\bibfnamefont {K.}~\bibnamefont {Kobayashi}}, \bibinfo
  {author} {\bibfnamefont {A.}~\bibnamefont {Fujimori}}, \bibinfo {author}
  {\bibfnamefont {T.}~\bibnamefont {Sasagawa}}, \bibinfo {author}
  {\bibfnamefont {T.}~\bibnamefont {Kimura}}, \bibinfo {author} {\bibfnamefont
  {K.}~\bibnamefont {Kishio}}, \bibinfo {author} {\bibfnamefont
  {K.}~\bibnamefont {Tamasaku}}, \bibinfo {author} {\bibfnamefont
  {H.}~\bibnamefont {Eisaki}}, \ and\ \bibinfo {author} {\bibfnamefont
  {S.}~\bibnamefont {Uchida}},\ }\href {\doibase 10.1103/PhysRevLett.81.2124}
  {\bibfield  {journal} {\bibinfo  {journal} {Phys. Rev. Lett.}\ }\textbf
  {\bibinfo {volume} {81}},\ \bibinfo {pages} {2124} (\bibinfo {year}
  {1998})}\BibitemShut {NoStop}%
\bibitem [{\citenamefont {Ronning}\ \emph {et~al.}(2003)\citenamefont
  {Ronning}, \citenamefont {Sasagawa}, \citenamefont {Kohsaka}, \citenamefont
  {Shen}, \citenamefont {Damascelli}, \citenamefont {Kim}, \citenamefont
  {Yoshida}, \citenamefont {Armitage}, \citenamefont {Lu}, \citenamefont
  {Feng}, \citenamefont {Miller}, \citenamefont {Takagi},\ and\ \citenamefont
  {Shen}}]{Ronning03}%
  \BibitemOpen
  \bibfield  {author} {\bibinfo {author} {\bibfnamefont {F.}~\bibnamefont
  {Ronning}}, \bibinfo {author} {\bibfnamefont {T.}~\bibnamefont {Sasagawa}},
  \bibinfo {author} {\bibfnamefont {Y.}~\bibnamefont {Kohsaka}}, \bibinfo
  {author} {\bibfnamefont {K.~M.}\ \bibnamefont {Shen}}, \bibinfo {author}
  {\bibfnamefont {A.}~\bibnamefont {Damascelli}}, \bibinfo {author}
  {\bibfnamefont {C.}~\bibnamefont {Kim}}, \bibinfo {author} {\bibfnamefont
  {T.}~\bibnamefont {Yoshida}}, \bibinfo {author} {\bibfnamefont {N.~P.}\
  \bibnamefont {Armitage}}, \bibinfo {author} {\bibfnamefont {D.~H.}\
  \bibnamefont {Lu}}, \bibinfo {author} {\bibfnamefont {D.~L.}\ \bibnamefont
  {Feng}}, \bibinfo {author} {\bibfnamefont {L.~L.}\ \bibnamefont {Miller}},
  \bibinfo {author} {\bibfnamefont {H.}~\bibnamefont {Takagi}}, \ and\ \bibinfo
  {author} {\bibfnamefont {Z.-X.}\ \bibnamefont {Shen}},\ }\href {\doibase
  10.1103/PhysRevB.67.165101} {\bibfield  {journal} {\bibinfo  {journal} {Phys.
  Rev. B}\ }\textbf {\bibinfo {volume} {67}},\ \bibinfo {pages} {165101}
  (\bibinfo {year} {2003})}\BibitemShut {NoStop}%
\bibitem [{\citenamefont {Fradkin}\ \emph {et~al.}(2015)\citenamefont
  {Fradkin}, \citenamefont {Kivelson},\ and\ \citenamefont
  {Tranquada}}]{Fradkin:2015ch}%
  \BibitemOpen
  \bibfield  {author} {\bibinfo {author} {\bibfnamefont {E.}~\bibnamefont
  {Fradkin}}, \bibinfo {author} {\bibfnamefont {S.~A.}\ \bibnamefont
  {Kivelson}}, \ and\ \bibinfo {author} {\bibfnamefont {J.~M.}\ \bibnamefont
  {Tranquada}},\ }\href {\doibase 10.1103/revmodphys.87.457} {\bibfield
  {journal} {\bibinfo  {journal} {Rev. Mod. Phys.}\ }\textbf {\bibinfo {volume}
  {87}},\ \bibinfo {pages} {457} (\bibinfo {year} {2015})}\BibitemShut
  {NoStop}%
\bibitem [{\citenamefont {P{\'e}pin}\ \emph {et~al.}(2020)\citenamefont
  {P{\'e}pin}, \citenamefont {Chakraborty}, \citenamefont {Grandadam},\ and\
  \citenamefont {Sarkar}}]{pepin2020fluctuations}%
  \BibitemOpen
  \bibfield  {author} {\bibinfo {author} {\bibfnamefont {C.}~\bibnamefont
  {P{\'e}pin}}, \bibinfo {author} {\bibfnamefont {D.}~\bibnamefont
  {Chakraborty}}, \bibinfo {author} {\bibfnamefont {M.}~\bibnamefont
  {Grandadam}}, \ and\ \bibinfo {author} {\bibfnamefont {S.}~\bibnamefont
  {Sarkar}},\ }\href@noop {} {\bibfield  {journal} {\bibinfo  {journal} {Annual
  Review of Condensed Matter Physics}\ }\textbf {\bibinfo {volume} {11}},\
  \bibinfo {pages} {301} (\bibinfo {year} {2020})}\BibitemShut {NoStop}%
\bibitem [{\citenamefont {Hoffman}\ \emph {et~al.}(2002)\citenamefont
  {Hoffman}, \citenamefont {Hudson}, \citenamefont {Lang}, \citenamefont
  {Madhavan}, \citenamefont {Eisaki}, \citenamefont {Uchida},\ and\
  \citenamefont {Davis}}]{Hoffman02}%
  \BibitemOpen
  \bibfield  {author} {\bibinfo {author} {\bibfnamefont {J.~E.}\ \bibnamefont
  {Hoffman}}, \bibinfo {author} {\bibfnamefont {E.~W.}\ \bibnamefont {Hudson}},
  \bibinfo {author} {\bibfnamefont {K.~M.}\ \bibnamefont {Lang}}, \bibinfo
  {author} {\bibfnamefont {V.}~\bibnamefont {Madhavan}}, \bibinfo {author}
  {\bibfnamefont {H.}~\bibnamefont {Eisaki}}, \bibinfo {author} {\bibfnamefont
  {S.}~\bibnamefont {Uchida}}, \ and\ \bibinfo {author} {\bibfnamefont {J.~C.}\
  \bibnamefont {Davis}},\ }\href {\doibase 10.1126/science.1066974} {\bibfield
  {journal} {\bibinfo  {journal} {Science}\ }\textbf {\bibinfo {volume}
  {295}},\ \bibinfo {pages} {466} (\bibinfo {year} {2002})}\BibitemShut
  {NoStop}%
\bibitem [{\citenamefont {Doiron-Leyraud}\ \emph {et~al.}(2007)\citenamefont
  {Doiron-Leyraud}, \citenamefont {Proust}, \citenamefont {LeBoeuf},
  \citenamefont {Levallois}, \citenamefont {Bonnemaison}, \citenamefont
  {Liang}, \citenamefont {Bonn}, \citenamefont {Hardy},\ and\ \citenamefont
  {Taillefer}}]{Doiron-Leyraud07}%
  \BibitemOpen
  \bibfield  {author} {\bibinfo {author} {\bibfnamefont {N.}~\bibnamefont
  {Doiron-Leyraud}}, \bibinfo {author} {\bibfnamefont {C.}~\bibnamefont
  {Proust}}, \bibinfo {author} {\bibfnamefont {D.}~\bibnamefont {LeBoeuf}},
  \bibinfo {author} {\bibfnamefont {J.}~\bibnamefont {Levallois}}, \bibinfo
  {author} {\bibfnamefont {J.-B.}\ \bibnamefont {Bonnemaison}}, \bibinfo
  {author} {\bibfnamefont {R.}~\bibnamefont {Liang}}, \bibinfo {author}
  {\bibfnamefont {D.~A.}\ \bibnamefont {Bonn}}, \bibinfo {author}
  {\bibfnamefont {W.~N.}\ \bibnamefont {Hardy}}, \ and\ \bibinfo {author}
  {\bibfnamefont {L.}~\bibnamefont {Taillefer}},\ }\href {\doibase
  10.1038/nature05872} {\bibfield  {journal} {\bibinfo  {journal} {Nature}\
  }\textbf {\bibinfo {volume} {447}},\ \bibinfo {pages} {565} (\bibinfo {year}
  {2007})}\BibitemShut {NoStop}%
\bibitem [{\citenamefont {Ghiringhelli}\ \emph {et~al.}(2012)\citenamefont
  {Ghiringhelli}, \citenamefont {Le~Tacon}, \citenamefont {Minola},
  \citenamefont {Blanco-Canosa}, \citenamefont {Mazzoli}, \citenamefont
  {Brookes}, \citenamefont {De~Luca}, \citenamefont {Frano}, \citenamefont
  {Hawthorn}, \citenamefont {He}, \citenamefont {Loew}, \citenamefont {Sala},
  \citenamefont {Peets}, \citenamefont {Salluzzo}, \citenamefont {Schierle},
  \citenamefont {Sutarto}, \citenamefont {Sawatzky}, \citenamefont {Weschke},
  \citenamefont {Keimer},\ and\ \citenamefont {Braicovich}}]{Ghiringhelli12}%
  \BibitemOpen
  \bibfield  {author} {\bibinfo {author} {\bibfnamefont {G.}~\bibnamefont
  {Ghiringhelli}}, \bibinfo {author} {\bibfnamefont {M.}~\bibnamefont
  {Le~Tacon}}, \bibinfo {author} {\bibfnamefont {M.}~\bibnamefont {Minola}},
  \bibinfo {author} {\bibfnamefont {S.}~\bibnamefont {Blanco-Canosa}}, \bibinfo
  {author} {\bibfnamefont {C.}~\bibnamefont {Mazzoli}}, \bibinfo {author}
  {\bibfnamefont {N.~B.}\ \bibnamefont {Brookes}}, \bibinfo {author}
  {\bibfnamefont {G.~M.}\ \bibnamefont {De~Luca}}, \bibinfo {author}
  {\bibfnamefont {A.}~\bibnamefont {Frano}}, \bibinfo {author} {\bibfnamefont
  {D.~G.}\ \bibnamefont {Hawthorn}}, \bibinfo {author} {\bibfnamefont
  {F.}~\bibnamefont {He}}, \bibinfo {author} {\bibfnamefont {T.}~\bibnamefont
  {Loew}}, \bibinfo {author} {\bibfnamefont {M.~M.}\ \bibnamefont {Sala}},
  \bibinfo {author} {\bibfnamefont {D.~C.}\ \bibnamefont {Peets}}, \bibinfo
  {author} {\bibfnamefont {M.}~\bibnamefont {Salluzzo}}, \bibinfo {author}
  {\bibfnamefont {E.}~\bibnamefont {Schierle}}, \bibinfo {author}
  {\bibfnamefont {R.}~\bibnamefont {Sutarto}}, \bibinfo {author} {\bibfnamefont
  {G.~A.}\ \bibnamefont {Sawatzky}}, \bibinfo {author} {\bibfnamefont
  {E.}~\bibnamefont {Weschke}}, \bibinfo {author} {\bibfnamefont
  {B.}~\bibnamefont {Keimer}}, \ and\ \bibinfo {author} {\bibfnamefont
  {L.}~\bibnamefont {Braicovich}},\ }\href {\doibase 10.1126/science.1223532}
  {\bibfield  {journal} {\bibinfo  {journal} {Science}\ }\textbf {\bibinfo
  {volume} {337}},\ \bibinfo {pages} {821} (\bibinfo {year}
  {2012})}\BibitemShut {NoStop}%
\bibitem [{\citenamefont {Wu}\ \emph {et~al.}(2012)\citenamefont {Wu},
  \citenamefont {Buchholz}, \citenamefont {Trabant}, \citenamefont {Chang},
  \citenamefont {Komarek}, \citenamefont {Heigl}, \citenamefont {Zimmermann},
  \citenamefont {Cwik}, \citenamefont {Nakamura}, \citenamefont {Braden} \emph
  {et~al.}}]{wu12}%
  \BibitemOpen
  \bibfield  {author} {\bibinfo {author} {\bibfnamefont {H.-H.}\ \bibnamefont
  {Wu}}, \bibinfo {author} {\bibfnamefont {M.}~\bibnamefont {Buchholz}},
  \bibinfo {author} {\bibfnamefont {C.}~\bibnamefont {Trabant}}, \bibinfo
  {author} {\bibfnamefont {C.}~\bibnamefont {Chang}}, \bibinfo {author}
  {\bibfnamefont {A.}~\bibnamefont {Komarek}}, \bibinfo {author} {\bibfnamefont
  {F.}~\bibnamefont {Heigl}}, \bibinfo {author} {\bibfnamefont
  {M.}~\bibnamefont {Zimmermann}}, \bibinfo {author} {\bibfnamefont
  {M.}~\bibnamefont {Cwik}}, \bibinfo {author} {\bibfnamefont {F.}~\bibnamefont
  {Nakamura}}, \bibinfo {author} {\bibfnamefont {M.}~\bibnamefont {Braden}},
  \emph {et~al.},\ }\href@noop {} {\bibfield  {journal} {\bibinfo  {journal}
  {Nature communications}\ }\textbf {\bibinfo {volume} {3}},\ \bibinfo {pages}
  {1023} (\bibinfo {year} {2012})}\BibitemShut {NoStop}%
\bibitem [{\citenamefont {Achkar}\ \emph {et~al.}(2012)\citenamefont {Achkar},
  \citenamefont {Sutarto}, \citenamefont {Mao}, \citenamefont {He},
  \citenamefont {Frano}, \citenamefont {Blanco-Canosa}, \citenamefont
  {Le~Tacon}, \citenamefont {Ghiringhelli}, \citenamefont {Braicovich},
  \citenamefont {Minola}, \citenamefont {Moretti~Sala}, \citenamefont
  {Mazzoli}, \citenamefont {Liang}, \citenamefont {Bonn}, \citenamefont
  {Hardy}, \citenamefont {Keimer}, \citenamefont {Sawatzky},\ and\
  \citenamefont {Hawthorn}}]{Achkar12}%
  \BibitemOpen
  \bibfield  {author} {\bibinfo {author} {\bibfnamefont {A.~J.}\ \bibnamefont
  {Achkar}}, \bibinfo {author} {\bibfnamefont {R.}~\bibnamefont {Sutarto}},
  \bibinfo {author} {\bibfnamefont {X.}~\bibnamefont {Mao}}, \bibinfo {author}
  {\bibfnamefont {F.}~\bibnamefont {He}}, \bibinfo {author} {\bibfnamefont
  {A.}~\bibnamefont {Frano}}, \bibinfo {author} {\bibfnamefont
  {S.}~\bibnamefont {Blanco-Canosa}}, \bibinfo {author} {\bibfnamefont
  {M.}~\bibnamefont {Le~Tacon}}, \bibinfo {author} {\bibfnamefont
  {G.}~\bibnamefont {Ghiringhelli}}, \bibinfo {author} {\bibfnamefont
  {L.}~\bibnamefont {Braicovich}}, \bibinfo {author} {\bibfnamefont
  {M.}~\bibnamefont {Minola}}, \bibinfo {author} {\bibfnamefont
  {M.}~\bibnamefont {Moretti~Sala}}, \bibinfo {author} {\bibfnamefont
  {C.}~\bibnamefont {Mazzoli}}, \bibinfo {author} {\bibfnamefont
  {R.}~\bibnamefont {Liang}}, \bibinfo {author} {\bibfnamefont {D.~A.}\
  \bibnamefont {Bonn}}, \bibinfo {author} {\bibfnamefont {W.~N.}\ \bibnamefont
  {Hardy}}, \bibinfo {author} {\bibfnamefont {B.}~\bibnamefont {Keimer}},
  \bibinfo {author} {\bibfnamefont {G.~A.}\ \bibnamefont {Sawatzky}}, \ and\
  \bibinfo {author} {\bibfnamefont {D.~G.}\ \bibnamefont {Hawthorn}},\ }\href
  {\doibase 10.1103/PhysRevLett.109.167001} {\bibfield  {journal} {\bibinfo
  {journal} {Phys. Rev. Lett.}\ }\textbf {\bibinfo {volume} {109}},\ \bibinfo
  {pages} {167001} (\bibinfo {year} {2012})}\BibitemShut {NoStop}%
\bibitem [{\citenamefont {Blackburn}\ \emph
  {et~al.}(2013{\natexlab{a}})\citenamefont {Blackburn}, \citenamefont {Chang},
  \citenamefont {H\"ucker}, \citenamefont {Holmes}, \citenamefont
  {Christensen}, \citenamefont {Liang}, \citenamefont {Bonn}, \citenamefont
  {Hardy}, \citenamefont {R\"utt}, \citenamefont {Gutowski}, \citenamefont
  {Zimmermann}, \citenamefont {Forgan},\ and\ \citenamefont
  {Hayden}}]{Blackburn13a}%
  \BibitemOpen
  \bibfield  {author} {\bibinfo {author} {\bibfnamefont {E.}~\bibnamefont
  {Blackburn}}, \bibinfo {author} {\bibfnamefont {J.}~\bibnamefont {Chang}},
  \bibinfo {author} {\bibfnamefont {M.}~\bibnamefont {H\"ucker}}, \bibinfo
  {author} {\bibfnamefont {A.~T.}\ \bibnamefont {Holmes}}, \bibinfo {author}
  {\bibfnamefont {N.~B.}\ \bibnamefont {Christensen}}, \bibinfo {author}
  {\bibfnamefont {R.}~\bibnamefont {Liang}}, \bibinfo {author} {\bibfnamefont
  {D.~A.}\ \bibnamefont {Bonn}}, \bibinfo {author} {\bibfnamefont {W.~N.}\
  \bibnamefont {Hardy}}, \bibinfo {author} {\bibfnamefont {U.}~\bibnamefont
  {R\"utt}}, \bibinfo {author} {\bibfnamefont {O.}~\bibnamefont {Gutowski}},
  \bibinfo {author} {\bibfnamefont {M.~v.}\ \bibnamefont {Zimmermann}},
  \bibinfo {author} {\bibfnamefont {E.~M.}\ \bibnamefont {Forgan}}, \ and\
  \bibinfo {author} {\bibfnamefont {S.~M.}\ \bibnamefont {Hayden}},\ }\href
  {\doibase 10.1103/PhysRevLett.110.137004} {\bibfield  {journal} {\bibinfo
  {journal} {Phys. Rev. Lett.}\ }\textbf {\bibinfo {volume} {110}},\ \bibinfo
  {pages} {137004} (\bibinfo {year} {2013}{\natexlab{a}})}\BibitemShut
  {NoStop}%
\bibitem [{\citenamefont {Blackburn}\ \emph
  {et~al.}(2013{\natexlab{b}})\citenamefont {Blackburn}, \citenamefont {Chang},
  \citenamefont {Said}, \citenamefont {Leu}, \citenamefont {Liang},
  \citenamefont {Bonn}, \citenamefont {Hardy}, \citenamefont {Forgan},\ and\
  \citenamefont {Hayden}}]{Blackburn13b}%
  \BibitemOpen
  \bibfield  {author} {\bibinfo {author} {\bibfnamefont {E.}~\bibnamefont
  {Blackburn}}, \bibinfo {author} {\bibfnamefont {J.}~\bibnamefont {Chang}},
  \bibinfo {author} {\bibfnamefont {A.~H.}\ \bibnamefont {Said}}, \bibinfo
  {author} {\bibfnamefont {B.~M.}\ \bibnamefont {Leu}}, \bibinfo {author}
  {\bibfnamefont {R.}~\bibnamefont {Liang}}, \bibinfo {author} {\bibfnamefont
  {D.~A.}\ \bibnamefont {Bonn}}, \bibinfo {author} {\bibfnamefont {W.~N.}\
  \bibnamefont {Hardy}}, \bibinfo {author} {\bibfnamefont {E.~M.}\ \bibnamefont
  {Forgan}}, \ and\ \bibinfo {author} {\bibfnamefont {S.~M.}\ \bibnamefont
  {Hayden}},\ }\href {\doibase 10.1103/PhysRevB.88.054506} {\bibfield
  {journal} {\bibinfo  {journal} {Phys. Rev. B}\ }\textbf {\bibinfo {volume}
  {88}},\ \bibinfo {pages} {054506} (\bibinfo {year}
  {2013}{\natexlab{b}})}\BibitemShut {NoStop}%
\bibitem [{\citenamefont {Blanco-Canosa}\ \emph {et~al.}(2013)\citenamefont
  {Blanco-Canosa}, \citenamefont {Frano}, \citenamefont {Loew}, \citenamefont
  {Lu}, \citenamefont {Porras}, \citenamefont {Ghiringhelli}, \citenamefont
  {Minola}, \citenamefont {Mazzoli}, \citenamefont {Braicovich}, \citenamefont
  {Schierle}, \citenamefont {Weschke}, \citenamefont {Le~Tacon},\ and\
  \citenamefont {Keimer}}]{Blanco-Canosa13}%
  \BibitemOpen
  \bibfield  {author} {\bibinfo {author} {\bibfnamefont {S.}~\bibnamefont
  {Blanco-Canosa}}, \bibinfo {author} {\bibfnamefont {A.}~\bibnamefont
  {Frano}}, \bibinfo {author} {\bibfnamefont {T.}~\bibnamefont {Loew}},
  \bibinfo {author} {\bibfnamefont {Y.}~\bibnamefont {Lu}}, \bibinfo {author}
  {\bibfnamefont {J.}~\bibnamefont {Porras}}, \bibinfo {author} {\bibfnamefont
  {G.}~\bibnamefont {Ghiringhelli}}, \bibinfo {author} {\bibfnamefont
  {M.}~\bibnamefont {Minola}}, \bibinfo {author} {\bibfnamefont
  {C.}~\bibnamefont {Mazzoli}}, \bibinfo {author} {\bibfnamefont
  {L.}~\bibnamefont {Braicovich}}, \bibinfo {author} {\bibfnamefont
  {E.}~\bibnamefont {Schierle}}, \bibinfo {author} {\bibfnamefont
  {E.}~\bibnamefont {Weschke}}, \bibinfo {author} {\bibfnamefont
  {M.}~\bibnamefont {Le~Tacon}}, \ and\ \bibinfo {author} {\bibfnamefont
  {B.}~\bibnamefont {Keimer}},\ }\href {\doibase
  10.1103/PhysRevLett.110.187001} {\bibfield  {journal} {\bibinfo  {journal}
  {Phys. Rev. Lett.}\ }\textbf {\bibinfo {volume} {110}},\ \bibinfo {pages}
  {187001} (\bibinfo {year} {2013})}\BibitemShut {NoStop}%
\bibitem [{\citenamefont {Croft}\ \emph {et~al.}(2014)\citenamefont {Croft},
  \citenamefont {Lester}, \citenamefont {Senn}, \citenamefont {Bombardi},\ and\
  \citenamefont {Hayden}}]{Croft14}%
  \BibitemOpen
  \bibfield  {author} {\bibinfo {author} {\bibfnamefont {T.~P.}\ \bibnamefont
  {Croft}}, \bibinfo {author} {\bibfnamefont {C.}~\bibnamefont {Lester}},
  \bibinfo {author} {\bibfnamefont {M.~S.}\ \bibnamefont {Senn}}, \bibinfo
  {author} {\bibfnamefont {A.}~\bibnamefont {Bombardi}}, \ and\ \bibinfo
  {author} {\bibfnamefont {S.~M.}\ \bibnamefont {Hayden}},\ }\href {\doibase
  10.1103/PhysRevB.89.224513} {\bibfield  {journal} {\bibinfo  {journal} {Phys.
  Rev. B}\ }\textbf {\bibinfo {volume} {89}},\ \bibinfo {pages} {224513}
  (\bibinfo {year} {2014})}\BibitemShut {NoStop}%
\bibitem [{\citenamefont {da~Silva~Neto}\ \emph {et~al.}(2014)\citenamefont
  {da~Silva~Neto}, \citenamefont {Aynajian}, \citenamefont {Frano},
  \citenamefont {Comin}, \citenamefont {Schierle}, \citenamefont {Weschke},
  \citenamefont {Gyenis}, \citenamefont {Wen}, \citenamefont {Schneeloch},
  \citenamefont {Xu}, \citenamefont {Ono}, \citenamefont {Gu}, \citenamefont
  {Le~Tacon},\ and\ \citenamefont {Yazdani}}]{daSilvaNeto:2014vy}%
  \BibitemOpen
  \bibfield  {author} {\bibinfo {author} {\bibfnamefont {E.~H.}\ \bibnamefont
  {da~Silva~Neto}}, \bibinfo {author} {\bibfnamefont {P.}~\bibnamefont
  {Aynajian}}, \bibinfo {author} {\bibfnamefont {A.}~\bibnamefont {Frano}},
  \bibinfo {author} {\bibfnamefont {R.}~\bibnamefont {Comin}}, \bibinfo
  {author} {\bibfnamefont {E.}~\bibnamefont {Schierle}}, \bibinfo {author}
  {\bibfnamefont {E.}~\bibnamefont {Weschke}}, \bibinfo {author} {\bibfnamefont
  {A.}~\bibnamefont {Gyenis}}, \bibinfo {author} {\bibfnamefont
  {J.}~\bibnamefont {Wen}}, \bibinfo {author} {\bibfnamefont {J.}~\bibnamefont
  {Schneeloch}}, \bibinfo {author} {\bibfnamefont {Z.}~\bibnamefont {Xu}},
  \bibinfo {author} {\bibfnamefont {S.}~\bibnamefont {Ono}}, \bibinfo {author}
  {\bibfnamefont {G.}~\bibnamefont {Gu}}, \bibinfo {author} {\bibfnamefont
  {M.}~\bibnamefont {Le~Tacon}}, \ and\ \bibinfo {author} {\bibfnamefont
  {A.}~\bibnamefont {Yazdani}},\ }\href
  {http://www.sciencemag.org/content/343/6169/393.abstract} {\bibfield
  {journal} {\bibinfo  {journal} {Science}\ }\textbf {\bibinfo {volume}
  {343}},\ \bibinfo {pages} {393} (\bibinfo {year} {2014})}\BibitemShut
  {NoStop}%
\bibitem [{\citenamefont {Matsuba}\ \emph {et~al.}(2007)\citenamefont
  {Matsuba}, \citenamefont {Yoshizawa}, \citenamefont {Mochizuki},
  \citenamefont {Mochiku}, \citenamefont {Hirata},\ and\ \citenamefont
  {Nishida}}]{matsuba07}%
  \BibitemOpen
  \bibfield  {author} {\bibinfo {author} {\bibfnamefont {K.}~\bibnamefont
  {Matsuba}}, \bibinfo {author} {\bibfnamefont {S.}~\bibnamefont {Yoshizawa}},
  \bibinfo {author} {\bibfnamefont {Y.}~\bibnamefont {Mochizuki}}, \bibinfo
  {author} {\bibfnamefont {T.}~\bibnamefont {Mochiku}}, \bibinfo {author}
  {\bibfnamefont {K.}~\bibnamefont {Hirata}}, \ and\ \bibinfo {author}
  {\bibfnamefont {N.}~\bibnamefont {Nishida}},\ }\href@noop {} {\bibfield
  {journal} {\bibinfo  {journal} {Journal of the Physical Society of Japan}\
  }\textbf {\bibinfo {volume} {76}},\ \bibinfo {pages} {063704} (\bibinfo
  {year} {2007})}\BibitemShut {NoStop}%
\bibitem [{\citenamefont {Fujita}\ \emph {et~al.}(2014)\citenamefont {Fujita},
  \citenamefont {Kim}, \citenamefont {Lee}, \citenamefont {Lee}, \citenamefont
  {Hamidian}, \citenamefont {Firmo}, \citenamefont {Mukhopadhyay},
  \citenamefont {Eisaki}, \citenamefont {Uchida}, \citenamefont {Lawler},
  \citenamefont {Kim},\ and\ \citenamefont {Davis}}]{Fujita14}%
  \BibitemOpen
  \bibfield  {author} {\bibinfo {author} {\bibfnamefont {K.}~\bibnamefont
  {Fujita}}, \bibinfo {author} {\bibfnamefont {C.~K.}\ \bibnamefont {Kim}},
  \bibinfo {author} {\bibfnamefont {I.}~\bibnamefont {Lee}}, \bibinfo {author}
  {\bibfnamefont {J.}~\bibnamefont {Lee}}, \bibinfo {author} {\bibfnamefont
  {M.}~\bibnamefont {Hamidian}}, \bibinfo {author} {\bibfnamefont {I.~A.}\
  \bibnamefont {Firmo}}, \bibinfo {author} {\bibfnamefont {S.}~\bibnamefont
  {Mukhopadhyay}}, \bibinfo {author} {\bibfnamefont {H.}~\bibnamefont
  {Eisaki}}, \bibinfo {author} {\bibfnamefont {S.}~\bibnamefont {Uchida}},
  \bibinfo {author} {\bibfnamefont {M.~J.}\ \bibnamefont {Lawler}}, \bibinfo
  {author} {\bibfnamefont {E.~A.}\ \bibnamefont {Kim}}, \ and\ \bibinfo
  {author} {\bibfnamefont {J.~C.}\ \bibnamefont {Davis}},\ }\href {\doibase
  10.1126/science.1248783} {\bibfield  {journal} {\bibinfo  {journal}
  {Science}\ }\textbf {\bibinfo {volume} {344}},\ \bibinfo {pages} {612}
  (\bibinfo {year} {2014})}\BibitemShut {NoStop}%
\bibitem [{\citenamefont {Machida}\ \emph {et~al.}(2016)\citenamefont
  {Machida}, \citenamefont {Kohsaka}, \citenamefont {Matsuoka}, \citenamefont
  {Iwaya}, \citenamefont {Hanaguri},\ and\ \citenamefont
  {Tamegai}}]{machida16}%
  \BibitemOpen
  \bibfield  {author} {\bibinfo {author} {\bibfnamefont {T.}~\bibnamefont
  {Machida}}, \bibinfo {author} {\bibfnamefont {Y.}~\bibnamefont {Kohsaka}},
  \bibinfo {author} {\bibfnamefont {K.}~\bibnamefont {Matsuoka}}, \bibinfo
  {author} {\bibfnamefont {K.}~\bibnamefont {Iwaya}}, \bibinfo {author}
  {\bibfnamefont {T.}~\bibnamefont {Hanaguri}}, \ and\ \bibinfo {author}
  {\bibfnamefont {T.}~\bibnamefont {Tamegai}},\ }\href@noop {} {\bibfield
  {journal} {\bibinfo  {journal} {Nature communications}\ }\textbf {\bibinfo
  {volume} {7}},\ \bibinfo {pages} {11747} (\bibinfo {year}
  {2016})}\BibitemShut {NoStop}%
\bibitem [{\citenamefont {Woll}\ and\ \citenamefont {Kohn}(1962)}]{Kohn1962}%
  \BibitemOpen
  \bibfield  {author} {\bibinfo {author} {\bibfnamefont {E.~J.}\ \bibnamefont
  {Woll}}\ and\ \bibinfo {author} {\bibfnamefont {W.}~\bibnamefont {Kohn}},\
  }\href {\doibase 10.1103/PhysRev.126.1693} {\bibfield  {journal} {\bibinfo
  {journal} {Phys. Rev.}\ }\textbf {\bibinfo {volume} {126}},\ \bibinfo {pages}
  {1693} (\bibinfo {year} {1962})}\BibitemShut {NoStop}%
\bibitem [{\citenamefont {Renker}\ \emph {et~al.}(1973)\citenamefont {Renker},
  \citenamefont {Rietschel}, \citenamefont {Pintschovius}, \citenamefont
  {Gl\"aser}, \citenamefont {Br\"uesch}, \citenamefont {Kuse},\ and\
  \citenamefont {Rice}}]{Renker1973}%
  \BibitemOpen
  \bibfield  {author} {\bibinfo {author} {\bibfnamefont {B.}~\bibnamefont
  {Renker}}, \bibinfo {author} {\bibfnamefont {H.}~\bibnamefont {Rietschel}},
  \bibinfo {author} {\bibfnamefont {L.}~\bibnamefont {Pintschovius}}, \bibinfo
  {author} {\bibfnamefont {W.}~\bibnamefont {Gl\"aser}}, \bibinfo {author}
  {\bibfnamefont {P.}~\bibnamefont {Br\"uesch}}, \bibinfo {author}
  {\bibfnamefont {D.}~\bibnamefont {Kuse}}, \ and\ \bibinfo {author}
  {\bibfnamefont {M.~J.}\ \bibnamefont {Rice}},\ }\href {\doibase
  10.1103/PhysRevLett.30.1144} {\bibfield  {journal} {\bibinfo  {journal}
  {Phys. Rev. Lett.}\ }\textbf {\bibinfo {volume} {30}},\ \bibinfo {pages}
  {1144} (\bibinfo {year} {1973})}\BibitemShut {NoStop}%
\bibitem [{\citenamefont {Carneiro}\ \emph {et~al.}(1976)\citenamefont
  {Carneiro}, \citenamefont {Shirane}, \citenamefont {Werner},\ and\
  \citenamefont {Kaiser}}]{Carneiro1976}%
  \BibitemOpen
  \bibfield  {author} {\bibinfo {author} {\bibfnamefont {K.}~\bibnamefont
  {Carneiro}}, \bibinfo {author} {\bibfnamefont {G.}~\bibnamefont {Shirane}},
  \bibinfo {author} {\bibfnamefont {S.~A.}\ \bibnamefont {Werner}}, \ and\
  \bibinfo {author} {\bibfnamefont {S.}~\bibnamefont {Kaiser}},\ }\href
  {\doibase 10.1103/PhysRevB.13.4258} {\bibfield  {journal} {\bibinfo
  {journal} {Phys. Rev. B}\ }\textbf {\bibinfo {volume} {13}},\ \bibinfo
  {pages} {4258} (\bibinfo {year} {1976})}\BibitemShut {NoStop}%
\bibitem [{\citenamefont {Pouget}\ \emph {et~al.}(1991)\citenamefont {Pouget},
  \citenamefont {Hennion}, \citenamefont {Escribe-Filippini},\ and\
  \citenamefont {Sato}}]{Pouget1991}%
  \BibitemOpen
  \bibfield  {author} {\bibinfo {author} {\bibfnamefont {J.~P.}\ \bibnamefont
  {Pouget}}, \bibinfo {author} {\bibfnamefont {B.}~\bibnamefont {Hennion}},
  \bibinfo {author} {\bibfnamefont {C.}~\bibnamefont {Escribe-Filippini}}, \
  and\ \bibinfo {author} {\bibfnamefont {M.}~\bibnamefont {Sato}},\ }\href
  {\doibase 10.1103/PhysRevB.43.8421} {\bibfield  {journal} {\bibinfo
  {journal} {Phys. Rev. B}\ }\textbf {\bibinfo {volume} {43}},\ \bibinfo
  {pages} {8421} (\bibinfo {year} {1991})}\BibitemShut {NoStop}%
\bibitem [{\citenamefont {Wilson}\ \emph {et~al.}(2001)\citenamefont {Wilson},
  \citenamefont {Di~Salvo},\ and\ \citenamefont {Mahajan}}]{wilson2001charge}%
  \BibitemOpen
  \bibfield  {author} {\bibinfo {author} {\bibfnamefont {J.}~\bibnamefont
  {Wilson}}, \bibinfo {author} {\bibfnamefont {F.}~\bibnamefont {Di~Salvo}}, \
  and\ \bibinfo {author} {\bibfnamefont {S.}~\bibnamefont {Mahajan}},\
  }\href@noop {} {\bibfield  {journal} {\bibinfo  {journal} {Advances in
  Physics}\ }\textbf {\bibinfo {volume} {50}},\ \bibinfo {pages} {1171}
  (\bibinfo {year} {2001})}\BibitemShut {NoStop}%
\bibitem [{\citenamefont {Le~Tacon}\ \emph {et~al.}(2014)\citenamefont
  {Le~Tacon}, \citenamefont {Bosak}, \citenamefont {Souliou}, \citenamefont
  {Dellea}, \citenamefont {Loew}, \citenamefont {Heid}, \citenamefont {Bohnen},
  \citenamefont {Ghiringhelli}, \citenamefont {Krisch},\ and\ \citenamefont
  {Keimer}}]{LeTacon14}%
  \BibitemOpen
  \bibfield  {author} {\bibinfo {author} {\bibfnamefont {M.}~\bibnamefont
  {Le~Tacon}}, \bibinfo {author} {\bibfnamefont {A.}~\bibnamefont {Bosak}},
  \bibinfo {author} {\bibfnamefont {S.~M.}\ \bibnamefont {Souliou}}, \bibinfo
  {author} {\bibfnamefont {G.}~\bibnamefont {Dellea}}, \bibinfo {author}
  {\bibfnamefont {T.}~\bibnamefont {Loew}}, \bibinfo {author} {\bibfnamefont
  {R.}~\bibnamefont {Heid}}, \bibinfo {author} {\bibfnamefont {K.-P.}\
  \bibnamefont {Bohnen}}, \bibinfo {author} {\bibfnamefont {G.}~\bibnamefont
  {Ghiringhelli}}, \bibinfo {author} {\bibfnamefont {M.}~\bibnamefont
  {Krisch}}, \ and\ \bibinfo {author} {\bibfnamefont {B.}~\bibnamefont
  {Keimer}},\ }\href {http://dx.doi.org/10.1038/nphys2805} {\bibfield
  {journal} {\bibinfo  {journal} {Nat. Phys.}\ }\textbf {\bibinfo {volume}
  {10}},\ \bibinfo {pages} {52} (\bibinfo {year} {2014})}\BibitemShut {NoStop}%
\bibitem [{\citenamefont {Miao}\ \emph {et~al.}(2018)\citenamefont {Miao},
  \citenamefont {Ishikawa}, \citenamefont {Heid}, \citenamefont {Le~Tacon},
  \citenamefont {Fabbris}, \citenamefont {Meyers}, \citenamefont {Gu},
  \citenamefont {Baron},\ and\ \citenamefont {Dean}}]{Miao18}%
  \BibitemOpen
  \bibfield  {author} {\bibinfo {author} {\bibfnamefont {H.}~\bibnamefont
  {Miao}}, \bibinfo {author} {\bibfnamefont {D.}~\bibnamefont {Ishikawa}},
  \bibinfo {author} {\bibfnamefont {R.}~\bibnamefont {Heid}}, \bibinfo {author}
  {\bibfnamefont {M.}~\bibnamefont {Le~Tacon}}, \bibinfo {author}
  {\bibfnamefont {G.}~\bibnamefont {Fabbris}}, \bibinfo {author} {\bibfnamefont
  {D.}~\bibnamefont {Meyers}}, \bibinfo {author} {\bibfnamefont {G.~D.}\
  \bibnamefont {Gu}}, \bibinfo {author} {\bibfnamefont {A.~Q.~R.}\ \bibnamefont
  {Baron}}, \ and\ \bibinfo {author} {\bibfnamefont {M.~P.~M.}\ \bibnamefont
  {Dean}},\ }\href {\doibase 10.1103/PhysRevX.8.011008} {\bibfield  {journal}
  {\bibinfo  {journal} {Phys. Rev. X}\ }\textbf {\bibinfo {volume} {8}},\
  \bibinfo {pages} {011008} (\bibinfo {year} {2018})}\BibitemShut {NoStop}%
\bibitem [{\citenamefont {Lee}\ \emph {et~al.}(2020)\citenamefont {Lee},
  \citenamefont {Zhou}, \citenamefont {Hepting}, \citenamefont {Li},
  \citenamefont {Nag}, \citenamefont {Walters}, \citenamefont
  {Garcia-Fernandez}, \citenamefont {Robarts}, \citenamefont {Hashimoto},
  \citenamefont {Lu} \emph {et~al.}}]{lee2020spectroscopic}%
  \BibitemOpen
  \bibfield  {author} {\bibinfo {author} {\bibfnamefont {W.}~\bibnamefont
  {Lee}}, \bibinfo {author} {\bibfnamefont {K.}~\bibnamefont {Zhou}}, \bibinfo
  {author} {\bibfnamefont {M.}~\bibnamefont {Hepting}}, \bibinfo {author}
  {\bibfnamefont {J.}~\bibnamefont {Li}}, \bibinfo {author} {\bibfnamefont
  {A.}~\bibnamefont {Nag}}, \bibinfo {author} {\bibfnamefont {A.}~\bibnamefont
  {Walters}}, \bibinfo {author} {\bibfnamefont {M.}~\bibnamefont
  {Garcia-Fernandez}}, \bibinfo {author} {\bibfnamefont {H.}~\bibnamefont
  {Robarts}}, \bibinfo {author} {\bibfnamefont {M.}~\bibnamefont {Hashimoto}},
  \bibinfo {author} {\bibfnamefont {H.}~\bibnamefont {Lu}},  \emph {et~al.},\
  }\href@noop {} {\bibfield  {journal} {\bibinfo  {journal} {arXiv preprint
  arXiv:2007.02464}\ } (\bibinfo {year} {2020})}\BibitemShut {NoStop}%
\bibitem [{\citenamefont {McQueeney}\ \emph {et~al.}(1999)\citenamefont
  {McQueeney}, \citenamefont {Petrov}, \citenamefont {Egami}, \citenamefont
  {Yethiraj}, \citenamefont {Shirane},\ and\ \citenamefont
  {Endoh}}]{Mcqueny1999}%
  \BibitemOpen
  \bibfield  {author} {\bibinfo {author} {\bibfnamefont {R.~J.}\ \bibnamefont
  {McQueeney}}, \bibinfo {author} {\bibfnamefont {Y.}~\bibnamefont {Petrov}},
  \bibinfo {author} {\bibfnamefont {T.}~\bibnamefont {Egami}}, \bibinfo
  {author} {\bibfnamefont {M.}~\bibnamefont {Yethiraj}}, \bibinfo {author}
  {\bibfnamefont {G.}~\bibnamefont {Shirane}}, \ and\ \bibinfo {author}
  {\bibfnamefont {Y.}~\bibnamefont {Endoh}},\ }\href {\doibase
  10.1103/PhysRevLett.82.628} {\bibfield  {journal} {\bibinfo  {journal} {Phys.
  Rev. Lett.}\ }\textbf {\bibinfo {volume} {82}},\ \bibinfo {pages} {628}
  (\bibinfo {year} {1999})}\BibitemShut {NoStop}%
\bibitem [{\citenamefont {Uchiyama}\ \emph {et~al.}(2004)\citenamefont
  {Uchiyama}, \citenamefont {Baron}, \citenamefont {Tsutsui}, \citenamefont
  {Tanaka}, \citenamefont {Hu}, \citenamefont {Yamamoto}, \citenamefont
  {Tajima},\ and\ \citenamefont {Endoh}}]{Uchiyama04}%
  \BibitemOpen
  \bibfield  {author} {\bibinfo {author} {\bibfnamefont {H.}~\bibnamefont
  {Uchiyama}}, \bibinfo {author} {\bibfnamefont {A.~Q.~R.}\ \bibnamefont
  {Baron}}, \bibinfo {author} {\bibfnamefont {S.}~\bibnamefont {Tsutsui}},
  \bibinfo {author} {\bibfnamefont {Y.}~\bibnamefont {Tanaka}}, \bibinfo
  {author} {\bibfnamefont {W.-Z.}\ \bibnamefont {Hu}}, \bibinfo {author}
  {\bibfnamefont {A.}~\bibnamefont {Yamamoto}}, \bibinfo {author}
  {\bibfnamefont {S.}~\bibnamefont {Tajima}}, \ and\ \bibinfo {author}
  {\bibfnamefont {Y.}~\bibnamefont {Endoh}},\ }\href {\doibase
  10.1103/PhysRevLett.92.197005} {\bibfield  {journal} {\bibinfo  {journal}
  {Phys. Rev. Lett.}\ }\textbf {\bibinfo {volume} {92}},\ \bibinfo {pages}
  {197005} (\bibinfo {year} {2004})}\BibitemShut {NoStop}%
\bibitem [{\citenamefont {Reznik}\ \emph {et~al.}(2008)\citenamefont {Reznik},
  \citenamefont {Fukuda}, \citenamefont {Lamago}, \citenamefont {Baron},
  \citenamefont {Tsutsui}, \citenamefont {Fujita},\ and\ \citenamefont
  {Yamada}}]{reznik2008q}%
  \BibitemOpen
  \bibfield  {author} {\bibinfo {author} {\bibfnamefont {D.}~\bibnamefont
  {Reznik}}, \bibinfo {author} {\bibfnamefont {T.}~\bibnamefont {Fukuda}},
  \bibinfo {author} {\bibfnamefont {D.}~\bibnamefont {Lamago}}, \bibinfo
  {author} {\bibfnamefont {A.}~\bibnamefont {Baron}}, \bibinfo {author}
  {\bibfnamefont {S.}~\bibnamefont {Tsutsui}}, \bibinfo {author} {\bibfnamefont
  {M.}~\bibnamefont {Fujita}}, \ and\ \bibinfo {author} {\bibfnamefont
  {K.}~\bibnamefont {Yamada}},\ }\href@noop {} {\bibfield  {journal} {\bibinfo
  {journal} {Journal of Physics and Chemistry of Solids}\ }\textbf {\bibinfo
  {volume} {69}},\ \bibinfo {pages} {3103} (\bibinfo {year}
  {2008})}\BibitemShut {NoStop}%
\bibitem [{\citenamefont {Graf}\ \emph {et~al.}(2008)\citenamefont {Graf},
  \citenamefont {d'Astuto}, \citenamefont {Jozwiak}, \citenamefont {Garcia},
  \citenamefont {Saini}, \citenamefont {Krisch}, \citenamefont {Ikeuchi},
  \citenamefont {Baron}, \citenamefont {Eisaki},\ and\ \citenamefont
  {Lanzara}}]{Graf08}%
  \BibitemOpen
  \bibfield  {author} {\bibinfo {author} {\bibfnamefont {J.}~\bibnamefont
  {Graf}}, \bibinfo {author} {\bibfnamefont {M.}~\bibnamefont {d'Astuto}},
  \bibinfo {author} {\bibfnamefont {C.}~\bibnamefont {Jozwiak}}, \bibinfo
  {author} {\bibfnamefont {D.~R.}\ \bibnamefont {Garcia}}, \bibinfo {author}
  {\bibfnamefont {N.~L.}\ \bibnamefont {Saini}}, \bibinfo {author}
  {\bibfnamefont {M.}~\bibnamefont {Krisch}}, \bibinfo {author} {\bibfnamefont
  {K.}~\bibnamefont {Ikeuchi}}, \bibinfo {author} {\bibfnamefont {A.~Q.~R.}\
  \bibnamefont {Baron}}, \bibinfo {author} {\bibfnamefont {H.}~\bibnamefont
  {Eisaki}}, \ and\ \bibinfo {author} {\bibfnamefont {A.}~\bibnamefont
  {Lanzara}},\ }\href {\doibase 10.1103/PhysRevLett.100.227002} {\bibfield
  {journal} {\bibinfo  {journal} {Phys. Rev. Lett.}\ }\textbf {\bibinfo
  {volume} {100}},\ \bibinfo {pages} {227002} (\bibinfo {year}
  {2008})}\BibitemShut {NoStop}%
\bibitem [{\citenamefont {d'Astuto}\ \emph {et~al.}(2008)\citenamefont
  {d'Astuto}, \citenamefont {Dhalenne}, \citenamefont {Graf}, \citenamefont
  {Hoesch}, \citenamefont {Giura}, \citenamefont {Krisch}, \citenamefont
  {Berthet}, \citenamefont {Lanzara},\ and\ \citenamefont {Shukla}}]{Astuto08}%
  \BibitemOpen
  \bibfield  {author} {\bibinfo {author} {\bibfnamefont {M.}~\bibnamefont
  {d'Astuto}}, \bibinfo {author} {\bibfnamefont {G.}~\bibnamefont {Dhalenne}},
  \bibinfo {author} {\bibfnamefont {J.}~\bibnamefont {Graf}}, \bibinfo {author}
  {\bibfnamefont {M.}~\bibnamefont {Hoesch}}, \bibinfo {author} {\bibfnamefont
  {P.}~\bibnamefont {Giura}}, \bibinfo {author} {\bibfnamefont
  {M.}~\bibnamefont {Krisch}}, \bibinfo {author} {\bibfnamefont
  {P.}~\bibnamefont {Berthet}}, \bibinfo {author} {\bibfnamefont
  {A.}~\bibnamefont {Lanzara}}, \ and\ \bibinfo {author} {\bibfnamefont
  {A.}~\bibnamefont {Shukla}},\ }\href {\doibase 10.1103/PhysRevB.78.140511}
  {\bibfield  {journal} {\bibinfo  {journal} {Phys. Rev. B}\ }\textbf {\bibinfo
  {volume} {78}},\ \bibinfo {pages} {140511} (\bibinfo {year}
  {2008})}\BibitemShut {NoStop}%
\bibitem [{\citenamefont {Baron}\ \emph {et~al.}(2008)\citenamefont {Baron},
  \citenamefont {Sutter}, \citenamefont {Tsutsui}, \citenamefont {Uchiyama},
  \citenamefont {Masui}, \citenamefont {Tajima}, \citenamefont {Heid},\ and\
  \citenamefont {Bohnen}}]{baron2008first}%
  \BibitemOpen
  \bibfield  {author} {\bibinfo {author} {\bibfnamefont {A.~Q.}\ \bibnamefont
  {Baron}}, \bibinfo {author} {\bibfnamefont {J.~P.}\ \bibnamefont {Sutter}},
  \bibinfo {author} {\bibfnamefont {S.}~\bibnamefont {Tsutsui}}, \bibinfo
  {author} {\bibfnamefont {H.}~\bibnamefont {Uchiyama}}, \bibinfo {author}
  {\bibfnamefont {T.}~\bibnamefont {Masui}}, \bibinfo {author} {\bibfnamefont
  {S.}~\bibnamefont {Tajima}}, \bibinfo {author} {\bibfnamefont
  {R.}~\bibnamefont {Heid}}, \ and\ \bibinfo {author} {\bibfnamefont {K.-P.}\
  \bibnamefont {Bohnen}},\ }\href@noop {} {\bibfield  {journal} {\bibinfo
  {journal} {Journal of Physics and Chemistry of Solids}\ }\textbf {\bibinfo
  {volume} {69}},\ \bibinfo {pages} {3100} (\bibinfo {year}
  {2008})}\BibitemShut {NoStop}%
\bibitem [{\citenamefont {Raichle}\ \emph {et~al.}(2011)\citenamefont
  {Raichle}, \citenamefont {Reznik}, \citenamefont {Lamago}, \citenamefont
  {Heid}, \citenamefont {Li}, \citenamefont {Bakr}, \citenamefont {Ulrich},
  \citenamefont {Hinkov}, \citenamefont {Hradil}, \citenamefont {Lin},\ and\
  \citenamefont {Keimer}}]{Raichle11}%
  \BibitemOpen
  \bibfield  {author} {\bibinfo {author} {\bibfnamefont {M.}~\bibnamefont
  {Raichle}}, \bibinfo {author} {\bibfnamefont {D.}~\bibnamefont {Reznik}},
  \bibinfo {author} {\bibfnamefont {D.}~\bibnamefont {Lamago}}, \bibinfo
  {author} {\bibfnamefont {R.}~\bibnamefont {Heid}}, \bibinfo {author}
  {\bibfnamefont {Y.}~\bibnamefont {Li}}, \bibinfo {author} {\bibfnamefont
  {M.}~\bibnamefont {Bakr}}, \bibinfo {author} {\bibfnamefont {C.}~\bibnamefont
  {Ulrich}}, \bibinfo {author} {\bibfnamefont {V.}~\bibnamefont {Hinkov}},
  \bibinfo {author} {\bibfnamefont {K.}~\bibnamefont {Hradil}}, \bibinfo
  {author} {\bibfnamefont {C.~T.}\ \bibnamefont {Lin}}, \ and\ \bibinfo
  {author} {\bibfnamefont {B.}~\bibnamefont {Keimer}},\ }\href {\doibase
  10.1103/PhysRevLett.107.177004} {\bibfield  {journal} {\bibinfo  {journal}
  {Phys. Rev. Lett.}\ }\textbf {\bibinfo {volume} {107}},\ \bibinfo {pages}
  {177004} (\bibinfo {year} {2011})}\BibitemShut {NoStop}%
\bibitem [{\citenamefont {Lorenzo}\ \emph {et~al.}(1998)\citenamefont
  {Lorenzo}, \citenamefont {Currat}, \citenamefont {Monceau}, \citenamefont
  {Hennion}, \citenamefont {Berger},\ and\ \citenamefont
  {Levy}}]{lorenzo1998neutron}%
  \BibitemOpen
  \bibfield  {author} {\bibinfo {author} {\bibfnamefont {J.}~\bibnamefont
  {Lorenzo}}, \bibinfo {author} {\bibfnamefont {R.}~\bibnamefont {Currat}},
  \bibinfo {author} {\bibfnamefont {P.}~\bibnamefont {Monceau}}, \bibinfo
  {author} {\bibfnamefont {B.}~\bibnamefont {Hennion}}, \bibinfo {author}
  {\bibfnamefont {H.}~\bibnamefont {Berger}}, \ and\ \bibinfo {author}
  {\bibfnamefont {F.}~\bibnamefont {Levy}},\ }\href@noop {} {\bibfield
  {journal} {\bibinfo  {journal} {Journal of Physics: Condensed Matter}\
  }\textbf {\bibinfo {volume} {10}},\ \bibinfo {pages} {5039} (\bibinfo {year}
  {1998})}\BibitemShut {NoStop}%
\bibitem [{\citenamefont {Requardt}\ \emph {et~al.}(2002)\citenamefont
  {Requardt}, \citenamefont {Lorenzo}, \citenamefont {Monceau}, \citenamefont
  {Currat},\ and\ \citenamefont {Krisch}}]{Requardt02}%
  \BibitemOpen
  \bibfield  {author} {\bibinfo {author} {\bibfnamefont {H.}~\bibnamefont
  {Requardt}}, \bibinfo {author} {\bibfnamefont {J.~E.}\ \bibnamefont
  {Lorenzo}}, \bibinfo {author} {\bibfnamefont {P.}~\bibnamefont {Monceau}},
  \bibinfo {author} {\bibfnamefont {R.}~\bibnamefont {Currat}}, \ and\ \bibinfo
  {author} {\bibfnamefont {M.}~\bibnamefont {Krisch}},\ }\href {\doibase
  10.1103/PhysRevB.66.214303} {\bibfield  {journal} {\bibinfo  {journal} {Phys.
  Rev. B}\ }\textbf {\bibinfo {volume} {66}},\ \bibinfo {pages} {214303}
  (\bibinfo {year} {2002})}\BibitemShut {NoStop}%
\bibitem [{\citenamefont {Efetov}\ \emph {et~al.}(2013)\citenamefont {Efetov},
  \citenamefont {Meier},\ and\ \citenamefont {P\'epin}}]{Efetov13}%
  \BibitemOpen
  \bibfield  {author} {\bibinfo {author} {\bibfnamefont {K.~B.}\ \bibnamefont
  {Efetov}}, \bibinfo {author} {\bibfnamefont {H.}~\bibnamefont {Meier}}, \
  and\ \bibinfo {author} {\bibfnamefont {C.}~\bibnamefont {P\'epin}},\ }\href
  {\doibase 10.1038/nphys2641} {\bibfield  {journal} {\bibinfo  {journal} {Nat.
  Phys.}\ }\textbf {\bibinfo {volume} {9}},\ \bibinfo {pages} {442} (\bibinfo
  {year} {2013})}\BibitemShut {NoStop}%
\bibitem [{\citenamefont {Hayward}\ \emph {et~al.}(2014)\citenamefont
  {Hayward}, \citenamefont {Hawthorn}, \citenamefont {Melko},\ and\
  \citenamefont {Sachdev}}]{Hayward14}%
  \BibitemOpen
  \bibfield  {author} {\bibinfo {author} {\bibfnamefont {L.~E.}\ \bibnamefont
  {Hayward}}, \bibinfo {author} {\bibfnamefont {D.~G.}\ \bibnamefont
  {Hawthorn}}, \bibinfo {author} {\bibfnamefont {R.~G.}\ \bibnamefont {Melko}},
  \ and\ \bibinfo {author} {\bibfnamefont {S.}~\bibnamefont {Sachdev}},\ }\href
  {\doibase 10.1126/science.1246310} {\bibfield  {journal} {\bibinfo  {journal}
  {Science}\ }\textbf {\bibinfo {volume} {343}},\ \bibinfo {pages} {1336}
  (\bibinfo {year} {2014})}\BibitemShut {NoStop}%
\bibitem [{\citenamefont {Wang}\ \emph {et~al.}(2015)\citenamefont {Wang},
  \citenamefont {Agterberg},\ and\ \citenamefont {Chubukov}}]{Wang15b}%
  \BibitemOpen
  \bibfield  {author} {\bibinfo {author} {\bibfnamefont {Y.}~\bibnamefont
  {Wang}}, \bibinfo {author} {\bibfnamefont {D.~F.}\ \bibnamefont {Agterberg}},
  \ and\ \bibinfo {author} {\bibfnamefont {A.}~\bibnamefont {Chubukov}},\
  }\href {\doibase 10.1103/PhysRevLett.114.197001} {\bibfield  {journal}
  {\bibinfo  {journal} {Phys. Rev. Lett.}\ }\textbf {\bibinfo {volume} {114}},\
  \bibinfo {pages} {197001} (\bibinfo {year} {2015})}\BibitemShut {NoStop}%
\bibitem [{\citenamefont {Chakraborty}\ \emph {et~al.}(2018)\citenamefont
  {Chakraborty}, \citenamefont {Morice},\ and\ \citenamefont
  {P{\'e}pin}}]{chakraborty2018phase}%
  \BibitemOpen
  \bibfield  {author} {\bibinfo {author} {\bibfnamefont {D.}~\bibnamefont
  {Chakraborty}}, \bibinfo {author} {\bibfnamefont {C.}~\bibnamefont {Morice}},
  \ and\ \bibinfo {author} {\bibfnamefont {C.}~\bibnamefont {P{\'e}pin}},\
  }\href@noop {} {\bibfield  {journal} {\bibinfo  {journal} {Physical Review
  B}\ }\textbf {\bibinfo {volume} {97}},\ \bibinfo {pages} {214501} (\bibinfo
  {year} {2018})}\BibitemShut {NoStop}%
\bibitem [{\citenamefont {Loret}\ \emph {et~al.}(2019)\citenamefont {Loret},
  \citenamefont {Auvray}, \citenamefont {Gallais}, \citenamefont {Cazayous},
  \citenamefont {Forget}, \citenamefont {Colson}, \citenamefont {Julien},
  \citenamefont {Paul}, \citenamefont {Civelli},\ and\ \citenamefont
  {Sacuto}}]{Loret19}%
  \BibitemOpen
  \bibfield  {author} {\bibinfo {author} {\bibfnamefont {B.}~\bibnamefont
  {Loret}}, \bibinfo {author} {\bibfnamefont {N.}~\bibnamefont {Auvray}},
  \bibinfo {author} {\bibfnamefont {Y.}~\bibnamefont {Gallais}}, \bibinfo
  {author} {\bibfnamefont {M.}~\bibnamefont {Cazayous}}, \bibinfo {author}
  {\bibfnamefont {A.}~\bibnamefont {Forget}}, \bibinfo {author} {\bibfnamefont
  {D.}~\bibnamefont {Colson}}, \bibinfo {author} {\bibfnamefont {M.-H.}\
  \bibnamefont {Julien}}, \bibinfo {author} {\bibfnamefont {I.}~\bibnamefont
  {Paul}}, \bibinfo {author} {\bibfnamefont {M.}~\bibnamefont {Civelli}}, \
  and\ \bibinfo {author} {\bibfnamefont {A.}~\bibnamefont {Sacuto}},\ }\href
  {https://www.nature.com/articles/s41567-019-0509-5} {\bibfield  {journal}
  {\bibinfo  {journal} {Nature Physics}\ ,\ \bibinfo {pages} {1}} (\bibinfo
  {year} {2019})}\BibitemShut {NoStop}%
\bibitem [{\citenamefont {Chang}\ \emph {et~al.}(2012)\citenamefont {Chang},
  \citenamefont {Blackburn}, \citenamefont {Holmes}, \citenamefont
  {Christensen}, \citenamefont {Larsen}, \citenamefont {Mesot}, \citenamefont
  {Liang}, \citenamefont {Bonn}, \citenamefont {Hardy}, \citenamefont
  {Watenphul}, \citenamefont {Zimmermann}, \citenamefont {Forgan},\ and\
  \citenamefont {Hayden}}]{Chang12}%
  \BibitemOpen
  \bibfield  {author} {\bibinfo {author} {\bibfnamefont {J.}~\bibnamefont
  {Chang}}, \bibinfo {author} {\bibfnamefont {E.}~\bibnamefont {Blackburn}},
  \bibinfo {author} {\bibfnamefont {A.~T.}\ \bibnamefont {Holmes}}, \bibinfo
  {author} {\bibfnamefont {N.~B.}\ \bibnamefont {Christensen}}, \bibinfo
  {author} {\bibfnamefont {J.}~\bibnamefont {Larsen}}, \bibinfo {author}
  {\bibfnamefont {J.}~\bibnamefont {Mesot}}, \bibinfo {author} {\bibfnamefont
  {R.}~\bibnamefont {Liang}}, \bibinfo {author} {\bibfnamefont {D.~A.}\
  \bibnamefont {Bonn}}, \bibinfo {author} {\bibfnamefont {W.~N.}\ \bibnamefont
  {Hardy}}, \bibinfo {author} {\bibfnamefont {A.}~\bibnamefont {Watenphul}},
  \bibinfo {author} {\bibfnamefont {M.~v.}\ \bibnamefont {Zimmermann}},
  \bibinfo {author} {\bibfnamefont {E.~M.}\ \bibnamefont {Forgan}}, \ and\
  \bibinfo {author} {\bibfnamefont {S.~M.}\ \bibnamefont {Hayden}},\ }\href
  {\doibase 10.1038/nphys2456} {\bibfield  {journal} {\bibinfo  {journal} {Nat.
  Phys.}\ }\textbf {\bibinfo {volume} {8}},\ \bibinfo {pages} {871} (\bibinfo
  {year} {2012})}\BibitemShut {NoStop}%
\bibitem [{\citenamefont {Chakraborty}\ \emph {et~al.}(2019)\citenamefont
  {Chakraborty}, \citenamefont {Grandadam}, \citenamefont {Hamidian},
  \citenamefont {Davis}, \citenamefont {Sidis},\ and\ \citenamefont
  {P\'epin}}]{Chakrabortyprb19}%
  \BibitemOpen
  \bibfield  {author} {\bibinfo {author} {\bibfnamefont {D.}~\bibnamefont
  {Chakraborty}}, \bibinfo {author} {\bibfnamefont {M.}~\bibnamefont
  {Grandadam}}, \bibinfo {author} {\bibfnamefont {M.~H.}\ \bibnamefont
  {Hamidian}}, \bibinfo {author} {\bibfnamefont {J.~C.~S.}\ \bibnamefont
  {Davis}}, \bibinfo {author} {\bibfnamefont {Y.}~\bibnamefont {Sidis}}, \ and\
  \bibinfo {author} {\bibfnamefont {C.}~\bibnamefont {P\'epin}},\ }\href
  {\doibase 10.1103/PhysRevB.100.224511} {\bibfield  {journal} {\bibinfo
  {journal} {Phys. Rev. B}\ }\textbf {\bibinfo {volume} {100}},\ \bibinfo
  {pages} {224511} (\bibinfo {year} {2019})}\BibitemShut {NoStop}%
\bibitem [{\citenamefont {Edkins}\ \emph {et~al.}(2019)\citenamefont {Edkins},
  \citenamefont {Kostin}, \citenamefont {Fujita}, \citenamefont {Mackenzie},
  \citenamefont {Eisaki}, \citenamefont {Uchida}, \citenamefont {Sachdev},
  \citenamefont {Lawler}, \citenamefont {Kim}, \citenamefont {Davis} \emph
  {et~al.}}]{edkins2019magnetic}%
  \BibitemOpen
  \bibfield  {author} {\bibinfo {author} {\bibfnamefont {S.~D.}\ \bibnamefont
  {Edkins}}, \bibinfo {author} {\bibfnamefont {A.}~\bibnamefont {Kostin}},
  \bibinfo {author} {\bibfnamefont {K.}~\bibnamefont {Fujita}}, \bibinfo
  {author} {\bibfnamefont {A.~P.}\ \bibnamefont {Mackenzie}}, \bibinfo {author}
  {\bibfnamefont {H.}~\bibnamefont {Eisaki}}, \bibinfo {author} {\bibfnamefont
  {S.}~\bibnamefont {Uchida}}, \bibinfo {author} {\bibfnamefont
  {S.}~\bibnamefont {Sachdev}}, \bibinfo {author} {\bibfnamefont {M.~J.}\
  \bibnamefont {Lawler}}, \bibinfo {author} {\bibfnamefont {E.-A.}\
  \bibnamefont {Kim}}, \bibinfo {author} {\bibfnamefont {J.~S.}\ \bibnamefont
  {Davis}},  \emph {et~al.},\ }\href@noop {} {\bibfield  {journal} {\bibinfo
  {journal} {Science}\ }\textbf {\bibinfo {volume} {364}},\ \bibinfo {pages}
  {976} (\bibinfo {year} {2019})}\BibitemShut {NoStop}%
\bibitem [{\citenamefont {Hamidian}\ \emph {et~al.}(2016)\citenamefont
  {Hamidian}, \citenamefont {Edkins}, \citenamefont {Joo}, \citenamefont
  {Kostin}, \citenamefont {Eisaki}, \citenamefont {Uchida}, \citenamefont
  {Lawler}, \citenamefont {Kim}, \citenamefont {Mackenzie}, \citenamefont
  {Fujita} \emph {et~al.}}]{hamidian2016detection}%
  \BibitemOpen
  \bibfield  {author} {\bibinfo {author} {\bibfnamefont {M.}~\bibnamefont
  {Hamidian}}, \bibinfo {author} {\bibfnamefont {S.}~\bibnamefont {Edkins}},
  \bibinfo {author} {\bibfnamefont {S.~H.}\ \bibnamefont {Joo}}, \bibinfo
  {author} {\bibfnamefont {A.}~\bibnamefont {Kostin}}, \bibinfo {author}
  {\bibfnamefont {H.}~\bibnamefont {Eisaki}}, \bibinfo {author} {\bibfnamefont
  {S.}~\bibnamefont {Uchida}}, \bibinfo {author} {\bibfnamefont
  {M.}~\bibnamefont {Lawler}}, \bibinfo {author} {\bibfnamefont {E.-A.}\
  \bibnamefont {Kim}}, \bibinfo {author} {\bibfnamefont {A.}~\bibnamefont
  {Mackenzie}}, \bibinfo {author} {\bibfnamefont {K.}~\bibnamefont {Fujita}},
  \emph {et~al.},\ }\href@noop {} {\bibfield  {journal} {\bibinfo  {journal}
  {Nature}\ }\textbf {\bibinfo {volume} {532}},\ \bibinfo {pages} {343}
  (\bibinfo {year} {2016})}\BibitemShut {NoStop}%
\bibitem [{\citenamefont {Norman}\ \emph
  {et~al.}(1998{\natexlab{a}})\citenamefont {Norman}, \citenamefont {Ding},
  \citenamefont {Randeria}, \citenamefont {Campuzano}, \citenamefont {Yokoya},
  \citenamefont {Takeuchi}, \citenamefont {Takahashi}, \citenamefont {Mochiku},
  \citenamefont {Kadowaki}, \citenamefont {Guptasarma} \emph
  {et~al.}}]{norman1998destruction}%
  \BibitemOpen
  \bibfield  {author} {\bibinfo {author} {\bibfnamefont {M.}~\bibnamefont
  {Norman}}, \bibinfo {author} {\bibfnamefont {H.}~\bibnamefont {Ding}},
  \bibinfo {author} {\bibfnamefont {M.}~\bibnamefont {Randeria}}, \bibinfo
  {author} {\bibfnamefont {J.}~\bibnamefont {Campuzano}}, \bibinfo {author}
  {\bibfnamefont {T.}~\bibnamefont {Yokoya}}, \bibinfo {author} {\bibfnamefont
  {T.}~\bibnamefont {Takeuchi}}, \bibinfo {author} {\bibfnamefont
  {T.}~\bibnamefont {Takahashi}}, \bibinfo {author} {\bibfnamefont
  {T.}~\bibnamefont {Mochiku}}, \bibinfo {author} {\bibfnamefont
  {K.}~\bibnamefont {Kadowaki}}, \bibinfo {author} {\bibfnamefont
  {P.}~\bibnamefont {Guptasarma}},  \emph {et~al.},\ }\href@noop {} {\bibfield
  {journal} {\bibinfo  {journal} {Nature}\ }\textbf {\bibinfo {volume} {392}},\
  \bibinfo {pages} {157} (\bibinfo {year} {1998}{\natexlab{a}})}\BibitemShut
  {NoStop}%
\bibitem [{\citenamefont {Norman}\ and\ \citenamefont
  {P\'epin}(2003)}]{Norman03}%
  \BibitemOpen
  \bibfield  {author} {\bibinfo {author} {\bibfnamefont {M.~R.}\ \bibnamefont
  {Norman}}\ and\ \bibinfo {author} {\bibfnamefont {C.}~\bibnamefont
  {P\'epin}},\ }\href {http://stacks.iop.org/0034-4885/66/i=10/a=R01}
  {\bibfield  {journal} {\bibinfo  {journal} {Rep. Prog. Phys.}\ }\textbf
  {\bibinfo {volume} {66}},\ \bibinfo {pages} {1547} (\bibinfo {year}
  {2003})}\BibitemShut {NoStop}%
\bibitem [{\citenamefont {Grandadam}\ \emph {et~al.}(2020)\citenamefont
  {Grandadam}, \citenamefont {Chakraborty}, \citenamefont {Montiel},\ and\
  \citenamefont {P{\'e}pin}}]{grandadam2020electronic}%
  \BibitemOpen
  \bibfield  {author} {\bibinfo {author} {\bibfnamefont {M.}~\bibnamefont
  {Grandadam}}, \bibinfo {author} {\bibfnamefont {D.}~\bibnamefont
  {Chakraborty}}, \bibinfo {author} {\bibfnamefont {X.}~\bibnamefont
  {Montiel}}, \ and\ \bibinfo {author} {\bibfnamefont {C.}~\bibnamefont
  {P{\'e}pin}},\ }\href@noop {} {\bibfield  {journal} {\bibinfo  {journal}
  {arXiv preprint arXiv:2002.12622}\ } (\bibinfo {year} {2020})}\BibitemShut
  {NoStop}%
\bibitem [{\citenamefont {Norman}\ \emph
  {et~al.}(1998{\natexlab{b}})\citenamefont {Norman}, \citenamefont {Randeria},
  \citenamefont {Ding},\ and\ \citenamefont {Campuzano}}]{Norman:1998va}%
  \BibitemOpen
  \bibfield  {author} {\bibinfo {author} {\bibfnamefont {M.~R.}\ \bibnamefont
  {Norman}}, \bibinfo {author} {\bibfnamefont {M.}~\bibnamefont {Randeria}},
  \bibinfo {author} {\bibfnamefont {H.}~\bibnamefont {Ding}}, \ and\ \bibinfo
  {author} {\bibfnamefont {J.~C.}\ \bibnamefont {Campuzano}},\ }\href {\doibase
  10.1103/PhysRevB.57.R11093} {\bibfield  {journal} {\bibinfo  {journal} {Phys.
  Rev. B}\ }\textbf {\bibinfo {volume} {57}},\ \bibinfo {pages} {R11093}
  (\bibinfo {year} {1998}{\natexlab{b}})}\BibitemShut {NoStop}%
\bibitem [{\citenamefont {Lee}\ \emph {et~al.}(1993)\citenamefont {Lee},
  \citenamefont {Rice},\ and\ \citenamefont {Anderson}}]{lee1993conductivity}%
  \BibitemOpen
  \bibfield  {author} {\bibinfo {author} {\bibfnamefont {P.}~\bibnamefont
  {Lee}}, \bibinfo {author} {\bibfnamefont {T.}~\bibnamefont {Rice}}, \ and\
  \bibinfo {author} {\bibfnamefont {P.}~\bibnamefont {Anderson}},\ }\href@noop
  {} {\bibfield  {journal} {\bibinfo  {journal} {Solid State Communications}\
  }\textbf {\bibinfo {volume} {88}},\ \bibinfo {pages} {1001} (\bibinfo {year}
  {1993})}\BibitemShut {NoStop}%
\bibitem [{\citenamefont {Berthod}\ \emph {et~al.}(2017)\citenamefont
  {Berthod}, \citenamefont {Maggio-Aprile}, \citenamefont {Bru\'er},
  \citenamefont {Erb},\ and\ \citenamefont {Renner}}]{Berthod17}%
  \BibitemOpen
  \bibfield  {author} {\bibinfo {author} {\bibfnamefont {C.}~\bibnamefont
  {Berthod}}, \bibinfo {author} {\bibfnamefont {I.}~\bibnamefont
  {Maggio-Aprile}}, \bibinfo {author} {\bibfnamefont {J.}~\bibnamefont
  {Bru\'er}}, \bibinfo {author} {\bibfnamefont {A.}~\bibnamefont {Erb}}, \ and\
  \bibinfo {author} {\bibfnamefont {C.}~\bibnamefont {Renner}},\ }\href
  {\doibase 10.1103/PhysRevLett.119.237001} {\bibfield  {journal} {\bibinfo
  {journal} {Phys. Rev. Lett.}\ }\textbf {\bibinfo {volume} {119}},\ \bibinfo
  {pages} {237001} (\bibinfo {year} {2017})}\BibitemShut {NoStop}%
\bibitem [{\citenamefont {Sarkar}\ \emph {et~al.}()\citenamefont {Sarkar},
  \citenamefont {Grandadam},\ and\ \citenamefont {P{\'e}pin}}]{SuppInf}%
  \BibitemOpen
  \bibfield  {author} {\bibinfo {author} {\bibfnamefont {S.}~\bibnamefont
  {Sarkar}}, \bibinfo {author} {\bibfnamefont {M.}~\bibnamefont {Grandadam}}, \
  and\ \bibinfo {author} {\bibfnamefont {C.}~\bibnamefont {P{\'e}pin}},\
  }\href@noop {} {\bibinfo  {journal} {Supplementary materials}\ }\BibitemShut
  {NoStop}%
\bibitem [{\citenamefont {Wang}\ and\ \citenamefont
  {Chubukov}(2014)}]{Wang:2014fr}%
  \BibitemOpen
\bibfield  {journal} {  }\bibfield  {author} {\bibinfo {author} {\bibfnamefont
  {Y.}~\bibnamefont {Wang}}\ and\ \bibinfo {author} {\bibfnamefont
  {A.}~\bibnamefont {Chubukov}},\ }\href@noop {} {\bibfield  {journal}
  {\bibinfo  {journal} {Physical Review B}\ }\textbf {\bibinfo {volume} {90}},\
  \bibinfo {pages} {2113} (\bibinfo {year} {2014})}\BibitemShut {NoStop}%
\bibitem [{\citenamefont {Chowdhury}\ and\ \citenamefont
  {Sachdev}(2014)}]{Chowdhury:2014cp}%
  \BibitemOpen
  \bibfield  {author} {\bibinfo {author} {\bibfnamefont {D.}~\bibnamefont
  {Chowdhury}}\ and\ \bibinfo {author} {\bibfnamefont {S.}~\bibnamefont
  {Sachdev}},\ }\href {\doibase 10.1103/PhysRevB.90.134516} {\bibfield
  {journal} {\bibinfo  {journal} {Phys. Rev. B}\ }\textbf {\bibinfo {volume}
  {90}},\ \bibinfo {pages} {134516} (\bibinfo {year} {2014})}\BibitemShut
  {NoStop}%
\bibitem [{\citenamefont {Comin}\ \emph {et~al.}(2014)\citenamefont {Comin},
  \citenamefont {Frano}, \citenamefont {Yee}, \citenamefont {Yoshida},
  \citenamefont {Eisaki}, \citenamefont {Schierle}, \citenamefont {Weschke},
  \citenamefont {Sutarto}, \citenamefont {He}, \citenamefont {Soumyanarayanan},
  \citenamefont {He}, \citenamefont {Le~Tacon}, \citenamefont {Elfimov},
  \citenamefont {Hoffman}, \citenamefont {Sawatzky}, \citenamefont {Keimer},\
  and\ \citenamefont {Damascelli}}]{Comin14}%
  \BibitemOpen
  \bibfield  {author} {\bibinfo {author} {\bibfnamefont {R.}~\bibnamefont
  {Comin}}, \bibinfo {author} {\bibfnamefont {A.}~\bibnamefont {Frano}},
  \bibinfo {author} {\bibfnamefont {M.~M.}\ \bibnamefont {Yee}}, \bibinfo
  {author} {\bibfnamefont {Y.}~\bibnamefont {Yoshida}}, \bibinfo {author}
  {\bibfnamefont {H.}~\bibnamefont {Eisaki}}, \bibinfo {author} {\bibfnamefont
  {E.}~\bibnamefont {Schierle}}, \bibinfo {author} {\bibfnamefont
  {E.}~\bibnamefont {Weschke}}, \bibinfo {author} {\bibfnamefont
  {R.}~\bibnamefont {Sutarto}}, \bibinfo {author} {\bibfnamefont
  {F.}~\bibnamefont {He}}, \bibinfo {author} {\bibfnamefont {A.}~\bibnamefont
  {Soumyanarayanan}}, \bibinfo {author} {\bibfnamefont {Y.}~\bibnamefont {He}},
  \bibinfo {author} {\bibfnamefont {M.}~\bibnamefont {Le~Tacon}}, \bibinfo
  {author} {\bibfnamefont {I.~S.}\ \bibnamefont {Elfimov}}, \bibinfo {author}
  {\bibfnamefont {J.~E.}\ \bibnamefont {Hoffman}}, \bibinfo {author}
  {\bibfnamefont {G.~A.}\ \bibnamefont {Sawatzky}}, \bibinfo {author}
  {\bibfnamefont {B.}~\bibnamefont {Keimer}}, \ and\ \bibinfo {author}
  {\bibfnamefont {A.}~\bibnamefont {Damascelli}},\ }\href {\doibase
  10.1126/science.1242996} {\bibfield  {journal} {\bibinfo  {journal}
  {Science}\ }\textbf {\bibinfo {volume} {343}},\ \bibinfo {pages} {390}
  (\bibinfo {year} {2014})}\BibitemShut {NoStop}%
\bibitem [{\citenamefont {Norman}\ \emph {et~al.}(1995)\citenamefont {Norman},
  \citenamefont {Randeria}, \citenamefont {Ding},\ and\ \citenamefont
  {Campuzano}}]{Norman:1995dd}%
  \BibitemOpen
  \bibfield  {author} {\bibinfo {author} {\bibfnamefont {M.~R.}\ \bibnamefont
  {Norman}}, \bibinfo {author} {\bibfnamefont {M.}~\bibnamefont {Randeria}},
  \bibinfo {author} {\bibfnamefont {H.}~\bibnamefont {Ding}}, \ and\ \bibinfo
  {author} {\bibfnamefont {J.~C.}\ \bibnamefont {Campuzano}},\ }\href {\doibase
  10.1103/physrevb.52.615} {\bibfield  {journal} {\bibinfo  {journal} {Phys.
  Rev. B}\ }\textbf {\bibinfo {volume} {52}},\ \bibinfo {pages} {615} (\bibinfo
  {year} {1995})}\BibitemShut {NoStop}%
\bibitem [{\citenamefont {Norman}\ \emph {et~al.}(2007)\citenamefont {Norman},
  \citenamefont {Kanigel}, \citenamefont {Randeria}, \citenamefont
  {Chatterjee},\ and\ \citenamefont {Campuzano}}]{NormanKanigel07}%
  \BibitemOpen
  \bibfield  {author} {\bibinfo {author} {\bibfnamefont {M.~R.}\ \bibnamefont
  {Norman}}, \bibinfo {author} {\bibfnamefont {A.}~\bibnamefont {Kanigel}},
  \bibinfo {author} {\bibfnamefont {M.}~\bibnamefont {Randeria}}, \bibinfo
  {author} {\bibfnamefont {U.}~\bibnamefont {Chatterjee}}, \ and\ \bibinfo
  {author} {\bibfnamefont {J.~C.}\ \bibnamefont {Campuzano}},\ }\href {\doibase
  10.1103/PhysRevB.76.174501} {\bibfield  {journal} {\bibinfo  {journal} {Phys.
  Rev. B}\ }\textbf {\bibinfo {volume} {76}},\ \bibinfo {pages} {174501}
  (\bibinfo {year} {2007})}\BibitemShut {NoStop}%
\bibitem [{\citenamefont {Dalla~Torre}\ \emph {et~al.}(2015)\citenamefont
  {Dalla~Torre}, \citenamefont {He}, \citenamefont {Benjamin},\ and\
  \citenamefont {Demler}}]{dalla2015exploring}%
  \BibitemOpen
  \bibfield  {author} {\bibinfo {author} {\bibfnamefont {E.~G.}\ \bibnamefont
  {Dalla~Torre}}, \bibinfo {author} {\bibfnamefont {Y.}~\bibnamefont {He}},
  \bibinfo {author} {\bibfnamefont {D.}~\bibnamefont {Benjamin}}, \ and\
  \bibinfo {author} {\bibfnamefont {E.}~\bibnamefont {Demler}},\ }\href@noop {}
  {\bibfield  {journal} {\bibinfo  {journal} {New Journal of Physics}\ }\textbf
  {\bibinfo {volume} {17}},\ \bibinfo {pages} {022001} (\bibinfo {year}
  {2015})}\BibitemShut {NoStop}%
\bibitem [{\citenamefont {Axe}\ and\ \citenamefont {Shirane}(1973)}]{Axe1973}%
  \BibitemOpen
  \bibfield  {author} {\bibinfo {author} {\bibfnamefont {J.~D.}\ \bibnamefont
  {Axe}}\ and\ \bibinfo {author} {\bibfnamefont {G.}~\bibnamefont {Shirane}},\
  }\href {\doibase 10.1103/PhysRevB.8.1965} {\bibfield  {journal} {\bibinfo
  {journal} {Phys. Rev. B}\ }\textbf {\bibinfo {volume} {8}},\ \bibinfo {pages}
  {1965} (\bibinfo {year} {1973})}\BibitemShut {NoStop}%
\bibitem [{\citenamefont {Reznik}(2012)}]{reznik2012phonon}%
  \BibitemOpen
  \bibfield  {author} {\bibinfo {author} {\bibfnamefont {D.}~\bibnamefont
  {Reznik}},\ }\href@noop {} {\bibfield  {journal} {\bibinfo  {journal}
  {Physica C: Superconductivity}\ }\textbf {\bibinfo {volume} {481}},\ \bibinfo
  {pages} {75} (\bibinfo {year} {2012})}\BibitemShut {NoStop}%
\bibitem [{\citenamefont {Chubukov}\ \emph {et~al.}(2007)\citenamefont
  {Chubukov}, \citenamefont {Norman}, \citenamefont {Millis},\ and\
  \citenamefont {Abrahams}}]{chubukov2007gapless}%
  \BibitemOpen
  \bibfield  {author} {\bibinfo {author} {\bibfnamefont {A.}~\bibnamefont
  {Chubukov}}, \bibinfo {author} {\bibfnamefont {M.}~\bibnamefont {Norman}},
  \bibinfo {author} {\bibfnamefont {A.}~\bibnamefont {Millis}}, \ and\ \bibinfo
  {author} {\bibfnamefont {E.}~\bibnamefont {Abrahams}},\ }\href@noop {}
  {\bibfield  {journal} {\bibinfo  {journal} {Physical Review B}\ }\textbf
  {\bibinfo {volume} {76}},\ \bibinfo {pages} {180501} (\bibinfo {year}
  {2007})}\BibitemShut {NoStop}%
\bibitem [{\citenamefont {Kanigel}\ \emph {et~al.}(2006)\citenamefont
  {Kanigel}, \citenamefont {Norman}, \citenamefont {Randeria}, \citenamefont
  {Chatterjee}, \citenamefont {Souma}, \citenamefont {Kaminski}, \citenamefont
  {Fretwell}, \citenamefont {Rosenkranz}, \citenamefont {Shi}, \citenamefont
  {Sato} \emph {et~al.}}]{kanigel2006evolution}%
  \BibitemOpen
  \bibfield  {author} {\bibinfo {author} {\bibfnamefont {A.}~\bibnamefont
  {Kanigel}}, \bibinfo {author} {\bibfnamefont {M.}~\bibnamefont {Norman}},
  \bibinfo {author} {\bibfnamefont {M.}~\bibnamefont {Randeria}}, \bibinfo
  {author} {\bibfnamefont {U.}~\bibnamefont {Chatterjee}}, \bibinfo {author}
  {\bibfnamefont {S.}~\bibnamefont {Souma}}, \bibinfo {author} {\bibfnamefont
  {A.}~\bibnamefont {Kaminski}}, \bibinfo {author} {\bibfnamefont
  {H.}~\bibnamefont {Fretwell}}, \bibinfo {author} {\bibfnamefont
  {S.}~\bibnamefont {Rosenkranz}}, \bibinfo {author} {\bibfnamefont
  {M.}~\bibnamefont {Shi}}, \bibinfo {author} {\bibfnamefont {T.}~\bibnamefont
  {Sato}},  \emph {et~al.},\ }\href@noop {} {\bibfield  {journal} {\bibinfo
  {journal} {Nature Physics}\ }\textbf {\bibinfo {volume} {2}},\ \bibinfo
  {pages} {447} (\bibinfo {year} {2006})}\BibitemShut {NoStop}%
\bibitem [{\citenamefont {Varma}\ \emph {et~al.}(1989)\citenamefont {Varma},
  \citenamefont {Littlewood}, \citenamefont {Schmitt-Rink}, \citenamefont
  {Abrahams},\ and\ \citenamefont {Ruckenstein}}]{Varma89}%
  \BibitemOpen
  \bibfield  {author} {\bibinfo {author} {\bibfnamefont {C.~M.}\ \bibnamefont
  {Varma}}, \bibinfo {author} {\bibfnamefont {P.~B.}\ \bibnamefont
  {Littlewood}}, \bibinfo {author} {\bibfnamefont {S.}~\bibnamefont
  {Schmitt-Rink}}, \bibinfo {author} {\bibfnamefont {E.}~\bibnamefont
  {Abrahams}}, \ and\ \bibinfo {author} {\bibfnamefont {A.~E.}\ \bibnamefont
  {Ruckenstein}},\ }\href {\doibase 10.1103/PhysRevLett.63.1996} {\bibfield
  {journal} {\bibinfo  {journal} {Phys. Rev. Lett.}\ }\textbf {\bibinfo
  {volume} {63}},\ \bibinfo {pages} {1996} (\bibinfo {year}
  {1989})}\BibitemShut {NoStop}%
\end{thebibliography}%
\end{document}


\setcounter{equation}{0}
\setcounter{figure}{0}
\setcounter{table}{0}
\setcounter{page}{1}
\makeatletter
\renewcommand{\theequation}{S\arabic{equation}}
\renewcommand{\thefigure}{S\arabic{figure}}
\title{Supplementary materials for\\ \vspace{5mm} Anomalous softening of phonon-dispersion in cuprate superconductors}
\author{Saheli Sarkar}
\author{Maxence Grandadam}
\author{Catherine P\'{e}pin}

\affiliation{Institut de Physique Th\'eorique, Universit\'e Paris-Saclay, CEA, CNRS, F-91191 Gif-sur-Yvette, France.}

\maketitle
\section{The model and parameters}
\begin{figure}[h]
\includegraphics[width=0.3\linewidth]{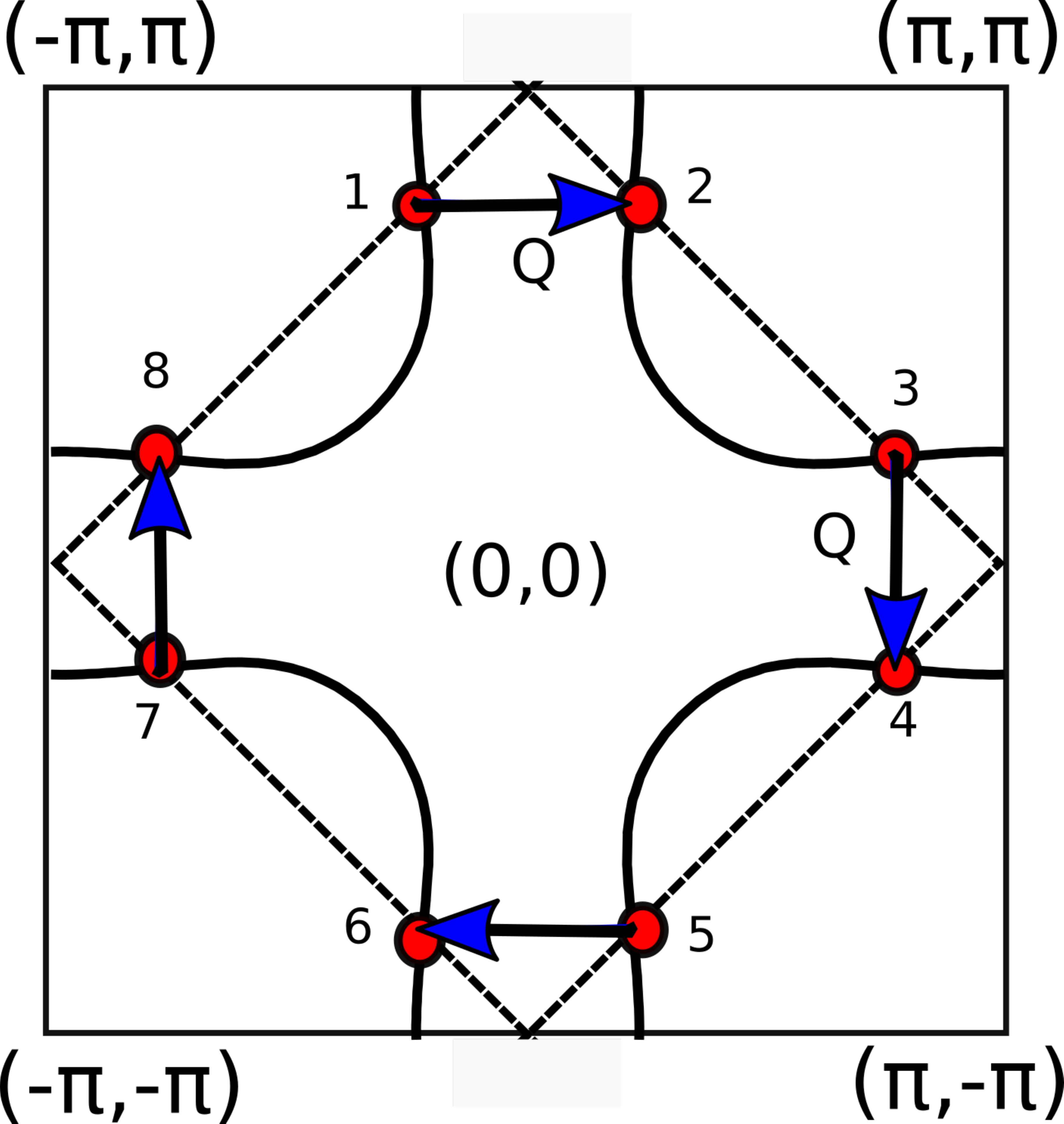}
\caption{\label{Fig_Fermisurface}Fermi surface for a prototype cuprate band structure. The solid black curves represent the Fermi-surface associated to the dispersion $\xi_{k}$. The dashed black lines represent the magnetic Brillouin zone boundary of the system. The intersection of the magnetic Brillouin zone boundary and the Fermi-surface are marked by red dots, representing the `hot-spots' on the Fermi-surface. The CDW modulation wave vector Q, indicated by the arrows are considered to be parallel to the crystallographic axes are shown in the figure. }
\end{figure}

In this section, we discuss the model describing the cuprate in the presence of charge-density wave (CDW) and superconducting (SC) orders. To analyze the phonon-dispersion in the presence of CDW and SC orders, we start with a total Hamiltonian $H_{tot}$, which incorporates an effective mean-field electronic Hamiltonian ($H_{e}$) describing CDW and SC orders, Hamiltonian for free phonons ($H_{ph}$) and electron-phonon interaction Hamiltonian ($H_{e-ph}$) as given in Eq.~(1) of the main text. The inverse Green's function matrix  $\hat{G}^{-1}(i\omega_{n},k) =( i \omega_{n} -\hat{H}_{e} )$ corresponding to the Hamiltonian $H_{e}$  in extended Nambu basis $\Psi^{\dagger}_{k} = \left(c^{\dagger}_{k,\uparrow},c_{-k,\downarrow},c^{\dagger}_{k+Q,\uparrow},c_{-k-Q,\downarrow}\right)$ is given by,
\begin{equation}\label{eq_invgreenmat}
G^{-1}(i\omega_{n},k)=
\begin{pmatrix}
i\omega_{n}-\xi_{k} & -\Delta_{k} & -\chi_{k} &0 \\
-\Delta^{*}_{k} & i\omega_{n}+\xi_{k} & 0 &\chi_{k}\\
-\chi^{*}_{k} &0 & i\omega_{n} - \xi_{k+Q} & -\Delta_{k+Q}\\
0 & \chi^{*}_{k} &-\Delta^{*}_{k+Q} & i\omega_{n}+\xi_{k+Q}
\end{pmatrix},
\end{equation}
where, $\xi_{k}$ is the electronic dispersion dispersion, given by, $\xi_{k} = 2 t_{1} [\cos k_{x}+\cos k_{y}] + 4 t_{2} \cos(k_{x})\cos(k_{y}) + 2 t_{3}( \cos(2 k_{x}) + \cos(2 k_{y}))-\mu$, with $t_{1}=-70.25$ meV, $t_{2} = 34.75$ meV, $t_{3} = -11$ meV and $\mu = -89$ meV. In this paper, all energy scales are expressed in units of $t_{1}$. $\Delta_{k}$ is the SC order parameter and $\chi_{k}$ is the CDW order parameter with finite wave-vector $Q$. $\omega_{n}$ is the Matsubara frequency. $\Psi$ is the Nambu spinor, $c^{\dagger}_{k,\uparrow}$ is the creation operator for an electronic state with wave-vector k and up spin and $c_{-k,\downarrow}$ is the annihilation operator for an electronic state with wave-vector -k and down spin. The Fermi-surface for the electronic dispersion $\xi_{k}$, `hot-spots' and the CDW wave vectors (Q) parallel to the crystallographic axes  are shown in Fig.~\ref{Fig_Fermisurface}. 
\section{ Dyson equations and calculation of the self-energy $\Pi$}
In this section ,we present the derivation of the phonon propagators, re-normalized due to CDW and SC orders.
The free phonon is given by the propagator $D_{0}(z,q)= 2 \omega_{q}/(z^{2}-\omega^{2}_{q})$, where $\omega_{q}$ is the frequency of the phonon mode for wave-vector $q$ and $z$ is a complex frequency (Im $z > 0$). The CDW and SC orders will couple to the free-phonon, modifying the propagator, which will consequently give rise to phonon modes with re-normalized dispersion. To evaluate the re-normalized dispersions, we start with Matsubara phonon propagator in matrix form whose elements are given by $D_{m,n}(q,\tau) = -\langle \mathcal{T} \phi_{q+mQ}(\tau)\phi^{\dagger}_{q+nQ}(0)\rangle$, where $\mathcal{T}$ is the time-ordering operator, and $m,n=\pm$. By noting that $D_{++} \equiv D_{--}:=D_{1}(z,q)$ and $D_{+-} \equiv D_{-+}:=D_{2}(z,q)$, within a perturbative approach for the electron-phonon interaction in Hamiltonian of Eq.~(1) in the main text, the Dyson's equations involving self-energies in the  presence of SC and CDW orders will give the modified phonon propagators $D_{1}$, $D_{2}$ and can be written as,
\begin{align}\label{eq:Suple_Dyson2}
D_{1}(z,q) & = D_{0}(z,q+Q)\bigg[1 + \Pi_{1}(z,q)D_{1}(z,q)+\Pi_{1}(z,q)D_{1}(z,q) +\Pi_{3}(z,q)D_{2}(z,q)+\Pi_{4}(z,q)D_{2}(z,q)\bigg],\\
\nonumber
 D_{2}(z,q) & =  D_{0}(z,q-Q)\bigg[\Pi_{5}(z,q)D_{2}(z,q)+\Pi_{6}(z,q)D_{2}(z,q) +\Pi_{7}(z,q)D_{1}(z,q)+ \Pi_{8}(z,q)D_{1}(z,q)\bigg].
\end{align}
The self-energies $\Pi_{1}, \Pi_{2}, \Pi_{3}$,  $\Pi_{4}$, $\Pi_{5}, \Pi_{6}, \Pi_{7}$ and $\Pi_{8}$ in Eq.~\ref{eq:Suple_Dyson2}, are given by,
\begin{align}\label{eq:suple_pi}
\Pi_{1}(\omega,q) & = \frac{g^{2}}{N}\sum_{k,i\omega_{n}}\left[G_{11}(k,i\omega_{n})G_{33}(k+q,i\omega_{n}+i\epsilon_{n}) +( k\rightarrow k-q)\right]\\
\nonumber
\Pi_{2}(\omega,q) & = \frac{g^{2}}{N}\sum_{k,i\omega_{n}}\left[G_{12}(k,i\omega_{n})G_{34}(k+q,i\omega_{n}+i\epsilon_{n}) +( k\rightarrow k-q)\right]\\
\nonumber
\Pi_{3}(\omega,q) & =\frac{g^{2}}{N} \sum_{k,i\omega_{n}}\left[G_{13}(k,i\omega_{n})G_{31}(k+q,i\omega_{n}+i\epsilon_{n}) +( k\rightarrow k-q)\right]\\
\nonumber
\Pi_{4}(\omega,q) & =\frac{g^{2}}{N} \sum_{k,i\omega_{n}}\left[G_{14}(k,i\omega_{n})G_{32}(k+q,i\omega_{n}+i\epsilon_{n}) +( k\rightarrow k-q)\right]\\
\nonumber
\Pi_{5}(\omega,q) & = \frac{g^{2}}{N}\sum_{k,i\omega_{n}}\left[G_{33}(k,i\omega_{n})G_{11}(k+q,i\omega_{n}+i\epsilon_{n}) +( k\rightarrow k-q)\right]\\
\nonumber
\Pi_{6}(\omega,q) & = \frac{g^{2}}{N}\sum_{k,i\omega_{n}}\left[G_{34}(k,i\omega_{n})G_{12}(k+q,i\omega_{n}+i\epsilon_{n}) +( k\rightarrow k-q)\right]\\
\nonumber
\Pi_{7}(\omega,q) & =\frac{g^{2}}{N} \sum_{k,i\omega_{n}}\left[G_{31}(k,i\omega_{n})G_{13}(k+q,i\omega_{n}+i\epsilon_{n}) +( k\rightarrow k-q)\right]\\
\nonumber
\Pi_{8}(\omega,q) & =\frac{g^{2}}{N} \sum_{k,i\omega_{n}}\left[G_{32}(k,i\omega_{n})G_{14}(k+q,i\omega_{n}+i\epsilon_{n}) +( k\rightarrow k-q)\right].
\end{align}
With a further assumption of small q, we note that $\Pi_{1}\approx \Pi_{5}$, $\Pi_{2}\approx \Pi_{6}$, $\Pi_{3}\approx \Pi_{7}$ and $\Pi_{4}\approx \Pi_{8}$, which gives the final form of the Dyson's equations as 
\begin{align}\label{eq:Suple_Dyson}
D_{1}(z,q) & = D_{0}(z,q+Q)\bigg[1 + \Pi_{1}(z,q)D_{1}(z,q)+\Pi_{2}(z,q)D_{1}(z,q) +\Pi_{3}(z,q)D_{2}(z,q)+\Pi_{4}(z,q)D_{2}(z,q)\bigg],\\
\nonumber
 D_{2}(z,q) & =  D_{0}(z,q-Q)\bigg[\Pi_{1}(z,q)D_{2}(z,q)+\Pi_{2}(z,q)D_{2}(z,q) +\Pi_{3}(z,q)D_{1}(z,q)+ \Pi_{4}(z,q)D_{1}(z,q)\bigg].
\end{align}
The corresponding Feynman-diagrams for the above self-energies in Eq.~\eqref{eq:Suple_Dyson} are shown in Fig.~(2) of the main text. In Eq.~\eqref{eq:Suple_Dyson}, we consider the strength of electron-phonon interaction, g to be $k$-independent and the number of lattice sites in the system to be N. 
The Dyson's equations from $D_{1}$ and $D_{2}$ can be decoupled to obtain the new re-normalized phonon modes, by introducing $D_{\pm}(z,q) = D_{1}(z,q) \pm D_{2}(z,q)$ and then solve for $D_{\pm}$.  The solution for the frequencies of the new phonon modes are,
\begin{align}\label{eq_SI:dispersion}
\Omega_{\pm}^{2}(q)=\omega_{Q}^{2}+2\omega_{Q}\Pi_{\pm}(q),
\end{align} 
where, $\Omega_{\pm}(q)$ represent the re-normalized frequencies [also given in Eq.~(5) of the main text], and $\Pi_{+} = \Pi_{1}+\Pi_{2}+\Pi_{3}+\Pi_{4} $ and $\Pi_{-} = \Pi_{1}+\Pi_{2}-\Pi_{3}-\Pi_{4}$.
The Green's function matrix elements [$G(i,j)$] that are appearing in the self-energy expressions in Eq.~\eqref{eq:Suple_Dyson}, are given by,
\begin{align}\label{eq:supp_greenmat}
G_{11}(k,i\omega_{n}) &= \frac{A_{1}}{(i\omega_{n}+E^{-}_{k})} + \frac{A_{2}}{(E^{-}_{k}-i\omega_{n})}+ \frac{A_{3}}{(E^{+}_{k}+i\omega_{n})}+\frac{A_{4}}{(E^{+}_{k}-i\omega_{n})},\\
\nonumber
G_{33}(k,i\omega_{n}) &= \frac{A_{5}}{(i\omega_{n}+E^{-}_{k})} + 
\frac{A_{6}}{(E^{-}_{k}-i\omega_{n})} + \frac{A_{7}}{(E^{+}_{k}+i\omega_{n})}+\frac{A_{8}}{(E^{+}_{k}-i\omega_{n})},\\
\nonumber
G_{12}(k,i\omega_{n}) &= \frac{A_{9}}{(i\omega_{n}+E^{-}_{k})} + \frac{A_{10}}{(E^{-}_{k}-i\omega_{n})} +\frac{A_{11}}{(E^{+}_{k}+i\omega_{n})}+\frac{A_{12}}{(E^{+}_{k}-i\omega_{n})},\\
\nonumber
G_{34}(k,i\omega_{n}) &= \frac{A_{13}}{(i\omega_{n}+E^{-}_{k})} + \frac{A_{14}}{(E^{-}_{k}-i\omega_{n})}+\frac{A_{15}}{(E^{+}_{k}+i\omega_{n})}+\frac{A_{16}}{(E^{+}_{k}-i\omega_{n})},\\
\nonumber
G_{13}(k,i\omega_{n}) &= \frac{A_{17}}{(i\omega_{n}+E^{-}_{k})} + \frac{A_{18}}{(E^{-}_{k}-i\omega_{n})}+ \frac{A_{19}}{(E^{+}_{k}+i\omega_{n})}+\frac{A_{20}}{(E^{+}_{k}-i\omega_{n})},\\
\nonumber
G_{31}(k,i\omega_{n}) &= \frac{A_{21}}{(i\omega_{n}+E^{-}_{k})} + \frac{A_{22}}{(E^{-}_{k}-i\omega_{n})}+ \frac{A_{23}}{(E^{+}_{k}+i\omega_{n})}+\frac{A_{24}}{(E^{+}_{k}-i\omega_{n})},\\
\nonumber
G_{14}(k,i\omega_{n}) &= \frac{A_{25}}{(i\omega_{n}+E^{-}_{k})} +\frac{A_{26}}{(E^{-}_{k}-i\omega_{n})}+ \frac{A_{27}}{(E^{+}_{k}+i\omega_{n})}+\frac{A_{28}}{(E^{+}_{k}-i\omega_{n})},\\
\nonumber
G_{32}(k,i\omega_{n}) &= \frac{A_{29}}{(i\omega_{n}+E^{-}_{k})} +\frac{A_{30}}{(E^{-}_{k}-i\omega_{n})}+ \frac{A_{31}}{(E^{+}_{k}+i\omega_{n})}+\frac{A_{32}}{(E^{+}_{k}-i\omega_{n})},
\end{align}
where, $E^{\pm}_{k}$ is the re-normalized electronic dispersion and is given by,
\begin{align}
E^{\pm}_{k} & = \pm\frac{1}{\sqrt{2}}\sqrt{\beta^{2}_{k}-\eta_{k}}
\end{align}
with,
\begin{align*}
\beta^{2}_{k} &= E^{2}_{1k} + E^{2}_{2k} + \Delta^{2}_{1k} + \Delta^{2}_{2k} + 2\chi^{2}_{k}, \\
\eta^{2}_{k} &=\left[(E_{1k}+E_{2k})(E_{1k}-E_{2k}) + (\Delta_{1k}+\Delta_{2k})(\Delta_{1k}-\Delta_{2k})\right]^{2}\\
& + 4\chi^{2}_{k}\left[(E_{1k}+E_{2k})^{2}+(\Delta_{1k}-\Delta_{2k})^{2}\right],
\end{align*}
where we use the following simplified notations, $ \xi_{k} = E_{1k} $, $\xi_{k+Q} = E_{2k}$, $\Delta_{k} = \Delta_{1k}$ and $\Delta_{k+Q} = \Delta_{2k}$. In the right hand side of the Eq.~\eqref{eq:supp_greenmat}, the numerators are given by,
\begin{align*}
A_{1} &= \frac{(E_{1k}-E^{-}_{k})(E^{2}_{2k}-(E^{-}_{k})^{2}+\Delta^{2}_{2k})-(E_{2k}+E^{-}_{k})\chi^{2}_{k}}{2E^{-}_{k}(E^{-}_{k}-E^{+}_{k})(E^{-}_{k}+E^{+}_{k})},\\
A_{2} &= \frac{(E_{1k}+E^{-}_{k})(E^{2}_{2k}-(E^{-}_{k})^{2}+\Delta^{2}_{2k})+(-E_{2k}+E^{-}_{k})\chi^{2}_{k}}{2E^{-}_{k}(E^{-}_{k}-E^{+}_{k})(E^{-}_{k}+E^{+}_{k})}\\
A_{3} &=\frac{-(E_{1k}-E^{+}_{k})(E^{2}_{2k}-(E^{+}_{k})^{2}+\Delta^{2}_{2k})+(E_{2k}+E^{+}_{k})\chi^{2}_{k}}{2E^{+}_{k}(E^{-}_{k}-E^{+}_{k})(E^{-}_{k}+E^{+}_{k})},\\
A_{4}& =\frac{-(E_{1k}+E^{+}_{k})(E^{2}_{2k}-(E^{+}_{k})^{2}+\Delta^{2}_{2k})+(E_{2k}-E^{+}_{k})\chi^{2}_{k}}{2E^{+}_{k}(E^{-}_{k}-E^{+}_{k})(E^{-}_{k}+E^{+}_{k})},
\end{align*}
\begin{align*}
A_{5} &= \frac{(E_{2k}-E^{-}_{k})(E^{2}_{1k}-(E^{-}_{k})^{2}+\Delta^{2}_{1k})-(E_{1k}+E^{-}_{k})\chi^{2}_{k}}{2E^{-}_{k}(E^{-}_{k}-E^{+}_{k})(E^{-}_{k}+E^{+}_{k})},\\
A_{6}& =\frac{(E_{2k}+E^{-}_{k})(E^{2}_{1k}-(E^{-}_{k})^{2}+\Delta^{2}_{1k})+(-E_{1k}+E^{-}_{k})\chi^{2}_{k}}{2E^{-}_{k}(E^{-}_{k}-E^{+}_{k})(E^{-}_{k}+E^{+}_{k})},\\
A_{7} &= \frac{-(E_{2k}-E^{+}_{k})(E^{2}_{1k}-(E^{+}_{k})^{2}+\Delta^{2}_{1k})+(E_{1k}+E^{+}_{k})\chi^{2}_{k}}{2E^{+}_{k}(E^{-}_{k}-E^{+}_{k})(E^{-}_{k}+E^{+}_{k})}\\
A_{8} &= \frac{-(E_{2k}+E^{+}_{k})(E^{2}_{1k}-(E^{+}_{k})^{2}+\Delta^{2}_{1k})+(E_{1k}-E^{+}_{k})\chi^{2}_{k}}{2E^{+}_{k}(E^{-}_{k}-E^{+}_{k})(E^{-}_{k}+E^{+}_{k})},
\end{align*}
\begin{align*}
A_{9} &=\frac{\Delta_{1k}(E^{2}_{2k}-(E^{-}_{k})^{2}+\Delta^{2}_{2k})+\Delta_{2k}\chi^{2}_{k}}{2E^{-}_{k}(E^{-}_{k}-E^{+}_{k})(E^{-}_{k}+E^{+}_{k})},\\
A_{10}& =\frac{\Delta_{1k}(E^{2}_{2k}-(E^{-}_{k})^{2}+\Delta^{2}_{2k})+\Delta_{2k}\chi^{2}_{k}}{2E^{-}_{k}(E^{-}_{k}-E^{+}_{k})(E^{-}_{k}+E^{+}_{k})},\\
A_{11} &=(-1) \frac{\Delta_{1k}(E^{2}_{2k}-(E^{+}_{k})^{2}+\Delta^{2}_{2k})+(\Delta_{2k})\chi^{2}_{k}}{2E^{+}_{k}(E^{-}_{k}-E^{+}_{k})(E^{-}_{k}+E^{+}_{k})},\\
A_{12} &=(-1) \frac{\Delta_{1k}(E^{2}_{2k}-(E^{+}_{k})^{2}+\Delta^{2}_{2k})+(\Delta_{2k})\chi^{2}_{k}}{2E^{+}_{k}(E^{-}_{k}-E^{+}_{k})(E^{-}_{k}+E^{+}_{k})},
\end{align*}
\begin{align*}
A_{13} &=\frac{\Delta_{2k}(E^{2}_{1k}-(E^{-}_{k})^{2}+\Delta^{2}_{1k})+\Delta_{1k}\chi^{2}_{k}}{2E^{-}_{k}(E^{-}_{k}-E^{+}_{k})(E^{-}_{k}+E^{+}_{k})},\\
A_{14} &=\frac{\Delta_{2k}(E^{2}_{1k}-(E^{-}_{k})^{2}+\Delta^{2}_{1k})+\Delta_{1k}\chi^{2}_{k}}{2E^{-}_{k}(E^{-}_{k}-E^{+}_{k})(E^{-}_{k}+E^{+}_{k})},\\
A_{15} &= (-1)\frac{\Delta_{2k}(E^{2}_{1k}-(E^{+}_{k})^{2}+\Delta^{2}_{1k})+(\Delta_{1k})\chi^{2}_{k}}{2E^{+}_{k}(E^{-}_{k}-E^{+}_{k})(E^{-}_{k}+E^{+}_{k})},\\
A_{16}& = (-1)\frac{\Delta_{2k}(E^{2}_{1k}-(E^{+}_{k})^{2}+\Delta^{2}_{1k})+(\Delta_{1k})\chi^{2}_{k}}{2E^{+}_{k}(E^{-}_{k}-E^{+}_{k})(E^{-}_{k}+E^{+}_{k})},
\end{align*}
\begin{align*}
A_{17}&= \frac{\chi_{k}[(E_{1k}-E^{-}_{k})(-E_{2k}+E^{-}_{k})+\Delta_{1k}\Delta_{2k}+\chi^{2}_{k}]}{2E^{-}_{k}(E^{-}_{k}-E^{+}_{k})(E^{-}_{k}+E^{+}_{k})},\\
A_{18} &=\frac{\chi_{k}[-(E_{1k}+E^{-}_{k})(E_{2k}+E^{-}_{k})+\Delta_{1k}\Delta_{2k}+\chi^{2}_{k}]}{2E^{-}_{k}(E^{-}_{k}-E^{+}_{k})(E^{-}_{k}+E^{+}_{k})},\\
A_{19} &=(-1)\frac{\chi_{k}[(E_{1k}-E^{+}_{k})(-E_{2k}+E^{+}_{k})+\Delta_{1k}\Delta_{2k}+\chi^{2}_{k}]}{2E^{+}_{k}(E^{-}_{k}-E^{+}_{k})(E^{-}_{k}+E^{+}_{k})},\\
A_{20} &= \frac{\chi_{k}[(E_{1k}+E^{+}_{k})(E_{2k}+E^{+}_{k})-\Delta_{1k}\Delta_{2k}-\chi^{2}_{k}]}{2E^{+}_{k}(E^{-}_{k}-E^{+}_{k})(E^{-}_{k}+E^{+}_{k})},
\end{align*}
\begin{align*}
\nonumber
A_{21}& = A_{17},\\
A_{22} &= A_{18},\\
A_{23} &= A_{19},\\
A_{24}& = A_{20},\\
A_{25} &= (-1)\frac{\chi_{k}[(E_{2k}+E^{-}_{k})(\Delta_{1k})+E_{1k}\Delta_{2k}-E^{-}_{k}\Delta_{2k}]}{2E^{-}_{k}(E^{-}_{k}-E^{+}_{k})(E^{-}_{k}+E^{+}_{k})},\\
A_{26} &= (-1)\frac{\chi_{k}[(E_{2k}-E^{-}_{k})(\Delta_{1k})+E_{1k}\Delta_{2k}+E^{-}_{k}\Delta_{2k}]}{2E^{-}_{k}(E^{-}_{k}-E^{+}_{k})(E^{-}_{k}+E^{+}_{k})},\\
A_{27} &= \frac{\chi_{k}[(E_{2k}+E^{+}_{k})(\Delta_{1k})+E_{1k}\Delta_{2k}-E^{+}_{k}\Delta_{2k}]}{2E^{+}_{k}(E^{-}_{k}-E^{+}_{k})(E^{-}_{k}+E^{+}_{k})},\\
A_{28}& = \frac{\chi_{k}[(E_{2k}-E^{+}_{k})(\Delta_{1k})+E_{1k}\Delta_{2k}+E^{+}_{k}\Delta_{2k}]}{2E^{+}_{k}(E^{-}_{k}-E^{+}_{k})(E^{-}_{k}+E^{+}_{k})},
\end{align*}
\begin{align*}
A_{29} &=(-1) \frac{\chi_{k}[(E_{2k}+E^{-}_{k})(\Delta_{1k})+E_{1k}\Delta_{2k}-E^{-}_{k}\Delta_{2k}]}{2E^{-}_{k}(E^{-}_{k}-E^{+}_{k})(E^{-}_{k}+E^{+}_{k})},\\
A_{30}& = (-1)\frac{\chi_{k}[(E_{2k}-E^{-}_{k})(\Delta_{1k})+E_{1k}\Delta_{2k}+E^{-}_{k}\Delta_{2k}]}{2E^{-}_{k}(E^{-}_{k}-E^{+}_{k})(E^{-}_{k}+E^{+}_{k})},\\
A_{31} &=\frac{\chi_{k}[(E_{2k}+E^{+}_{k})(\Delta_{1k})+E_{1k}\Delta_{2k}-E^{+}_{k}\Delta_{2k}]}{2E^{+}_{k}(E^{-}_{k}-E^{+}_{k})(E^{-}_{k}+E^{+}_{k})},\\
A_{32} &= \frac{\chi_{k}[(E_{2k}-E^{+}_{k})(\Delta_{1k})+E_{1k}\Delta_{2k}+E^{+}_{k}\Delta_{2k}]}{2E^{+}_{k}(E^{-}_{k}-E^{+}_{k})(E^{-}_{k}+E^{+}_{k})}.
\end{align*}
To evaluate the self-energies ($\Pi$) in Eq.~\eqref{eq:suple_pi}, we first perform the summation over the Matsubara frequency using analytic tool of contour integration. Next, we do the summation over $k$ using numerical tools taking $\Delta_{k} = \Delta_{k+Q}$, $g = 1$ and $N = 40000$. The plot of the self-energy in this case is presented in Fig.~(3) in the main text.
\section{Calculation of self-energy $\Pi$ in the presence of inverse life-time $\Gamma$}
\begin{figure}[h]
\includegraphics[width=0.4\linewidth]{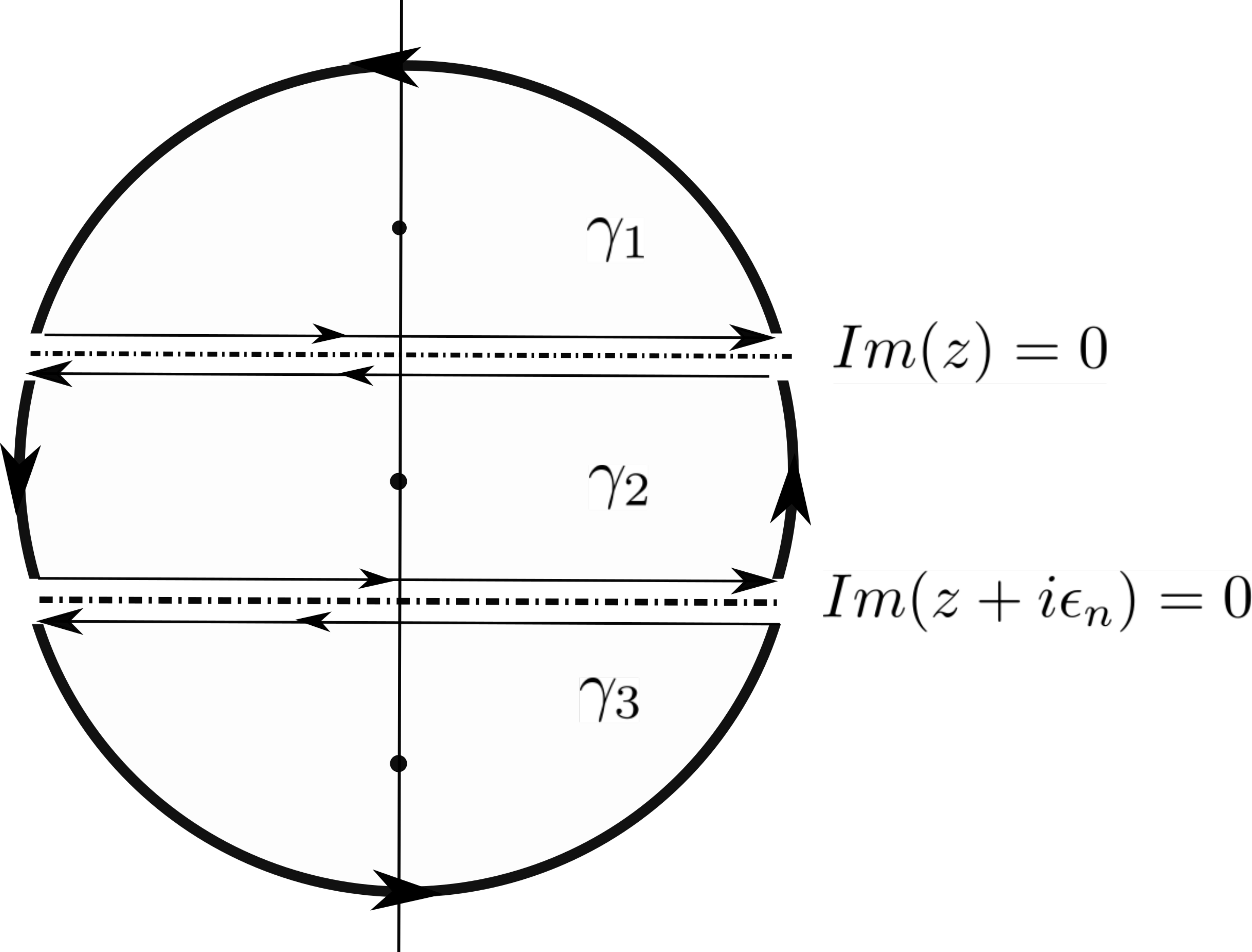}
\caption{\label{fig:contour} Contour for complex Matsubara frequency summation: $Im (z) = 0$ and $Im (z + i \epsilon_{n}) = 0$ denote the two branch-cuts in the complex plane. $\gamma_{1}$, $\gamma_{2}$ and $\gamma_{3}$ are the three contours of integration.}
\end{figure}

In this section, we present the self-energy calculation in the presence of finite inverse life-time ($\Gamma$) of the quasi-particles, associated to thermal fluctuations. Here, Green's function elements become,
\begin{align*}
G_{i,j}(i\omega_{n},k) \rightarrow  G_{i,j}(i\omega_{n}+i\Gamma sgn(\omega_{n}),k).
\end{align*} 
The self-energies $\Pi$ in Eq.~\eqref{eq:suple_pi} now have the following general structure:
\begin{align}
\sum_{k, i \omega_{n}}G^{a}_{k}(i \omega_{n} + i\Gamma sgn(\omega_{n}))G^{b}_{k+q}(i \omega_{n}+ i\Gamma sgn(\omega_{n}) + i \epsilon_{n}),
\end{align}
where, either of `a' and `b' symbolically represents the (i, j)$^{th}$ element of the Green's function matrix. To evaluate the Matsubara summation in this case, we need to use a contour avoiding the branch cuts defined by $Im (z) = 0$ and $Im (z + i\epsilon_{n}) = 0$ as shown in the Fig.~\ref{fig:contour}. Using this contour, we arrive at the following integrations,
\begin{align}\label{eq:supl_contour_I}
\begin{split}
I_{\gamma_{1}} &=\sum_{k}\left[\int_{-\infty}^{\infty} \frac{d\omega}{2\pi i}n_{F}(\omega) G^{a}_{k}(\omega + i \Gamma)G^{b}_{k+q}(\omega + \epsilon + i \Gamma) \right], \\
I_{\gamma_{3}} &=-\sum_{k}\left[\int_{-\infty}^{\infty} \frac{d\omega}{2\pi i}n_{F}(\omega) G^{a}_{k}(\omega -\epsilon - i \Gamma)G^{b}_{k+q}(\omega - i \Gamma)\right] ,\\
I_{\gamma_{2}} &=\sum_{k}\left[\int_{-\infty}^{\infty} \frac{d\omega}{2\pi i}n_{F}(\omega)\left\lbrace G^{a}_{k}(\omega -\epsilon- i \Gamma)G^{b}_{k+q}(\omega + i \Gamma) - G^{a}_{k}(\omega - i \Gamma)G^{b}_{k+q}(\omega + \epsilon + i \Gamma)\right\rbrace \right] ,
\end{split}
\end{align}
where, $n_{F}(\omega) = 1/(\exp \beta \omega + 1)$ is the Fermi-distribution function. Moreover $\beta = 1/k_{B}T$, where $k_{B}$ is the Boltzmann constant.

Next, in the limit $T \rightarrow 0$, the integrals $I_{\gamma_{1}}$ and $I_{\gamma_{3}}$ in Eq.~\eqref{eq:supl_contour_I} become,
\begin{align}\label{eq:supl_contour_13}
\begin{split}
I_{\gamma_{1}} &=\sum_{k}\left[\int_{-\infty}^{0} \frac{d\omega}{2\pi i} G^{a}_{k}(\omega + i \Gamma)G^{b}_{k+q}(\omega + \epsilon + i \Gamma) \right] \\
I_{\gamma_{3}} &=-\sum_{k}\left[\int_{-\infty}^{0} \frac{d\omega}{2\pi i} G^{a}_{k}(\omega -\epsilon - i \Gamma)G^{b}_{k+q}(\omega - i \Gamma)\right] .
\end{split}
\end{align}
We replace $\omega\rightarrow \omega + \epsilon$ in the first term of $I_{\gamma_{2}}$ and successively use
\begin{align*}
\lim_{\epsilon \rightarrow 0} \frac{n_{F}(\omega + \epsilon)- n_{F}(\omega)}{\epsilon} &= - \delta(\omega),
\end{align*}
where $\delta(\omega)$ is a Dirac delta function with the property, $\int_{-\infty}^{\infty} d\omega f(\omega)\delta(\omega) = f(0)$. Thus, $I_{\gamma_{2}}$ in Eq.~\eqref{eq:supl_contour_I} becomes,
\begin{align}\label{eq:supl_contour2}
I_{\gamma_{2}} &= \sum_{k}\frac{-\epsilon}{2\pi i}\left[G^{a}_{k}(i\Gamma)G^{b}_{k+q}(\epsilon +i \Gamma)\right].
\end{align}
Finally, we evaluate the real frequency ($\omega$) integration in Eq.~\eqref{eq:supl_contour_13} for each of the four self-energies $\Pi_{1}$, $\Pi_{2}$, $\Pi_{3}$ and $\Pi_{4}$ in Eq.~\eqref{eq:suple_pi} by using,
\begin{align}\label{int_omega}
\int_{-\infty}^{0}d\omega\left[\frac{1}{(\omega - x)*(\omega - y)}\right]&= \frac{\log [x] - \log [y]}{x-y},
\end{align}
where, $x,y \in \mathbb{C}$. The summation over $k$ is again evaluated using numerical tools. The re-normalization of phonon-spectrum in this case is given by the real part of the self-energy. The real part of the self-energy are plotted for different set of SC order, $\Gamma$ and CDW order in Fig.~(4) and Fig.~(5) of the main text.